# Josephson diode effect derived from short-range coherent coupling

Sadashige Matsuo[1], Takaya Imoto[1,2], Tomohiro Yokoyama[3], Yosuke Sato[1], Tyler Lindemann[4], Sergei Gronin[4], Geoffrey C. Gardner[4], Michael J. Manfra[4,5,6,7], Seigo Tarucha[1,8]

[1] *Center for Emergent Matter Science, RIKEN, Saitama 351-0198, Japan*
[2] *Department of Physics, Tokyo University of Science, Tokyo 162-8601, Japan*
[3] *Department of Material Engineering Science, Graduate School of Engineering Science, Osaka University,1-3 Machikaneyama, Toyonaka, Osaka, Japan*
[4] *Birck Nanotechnology Center, Purdue University, Indiana 47907, USA*
[5] *Department of Physics and Astronomy, Purdue University, Indiana 47907, USA*
[6] *School of Materials Engineering, Purdue University, Indiana 47907, USA*
[7] *School of Electrical and Computer Engineering, Purdue University, Indiana 47907, USA*
[8] *RIKEN Center for Quantum Computing, RIKEN, Saitama 351-0198, Japan*

**Superconducting devices with broken time-reversal and spatial-inversion symmetries can exhibit novel superconducting phenomena. The observation of superconducting diode effects, which is applicable for dissipationless rectification, provides information on the breaking of such symmetries. We experimentally study a Josephson junction (JJ) coupled to another adjacent JJ as a new system exhibiting the superconducting diode effect. We demonstrate that the observed superconducting diode effect can be controlled non-locally based on the phase difference with the adjacent JJ. These results indicate that the time-reversal and spatial-inversion symmetries of a JJ are broken by the coherent coupling to an adjacent JJ, and this enables the engineering of novel superconducting phenomena mediated by coherent coupling among JJs and development of their applications for superconducting diode devices.**

**Introduction**

  Symmetry breaking in superconducting (SC) devices has resulted in the emergence of exotic phenomena. For example, the emergence of Majorana zero modes in SC states with broken time-reversal and spatial-inversion symmetries has recently been proposed, and possible experimental signatures have been reported [1]. This implies that the analysis and control of the symmetry-breaking mechanism in SC devices are crucial not only for understanding such novel SC phenomena but also for providing flexibility in their engineering design.

  Superconducting diode effect (SDE), which is a direction-dependent supercurrent, namely, the nonreciprocal SC transport in SC devices, is a characteristic phenomenon occasionally found in SC states with both time-reversal and spatial-inversion symmetry breaking. Nonreciprocal transport can be employed for the rectification of electrical currents represented by p–n junctions. Therefore, dissipationless rectification can be achieved based on SDE. Several mechanisms generating SDE have been demonstrated in Josephson junction (JJ) devices [2–11], such as in asymmetric SC quantum interference devices (SQUIDs), where the time-reversal and spatial-inversion symmetries are broken by the induced phase difference and two asymmetric JJs, respectively [12,13]. Recently, SDE has also been observed in non-centrosymmetric bulk materials under external magnetic fields or even zero magnetic field [14–19].

In this study, we focus on two JJs with short-range coherent coupling [20–23]. This short-range coherent coupling is defined as the hybridization of the Andreev bound states in the respective JJs through the tunneling effect, as described schematically in Fig. 1(a). In this device, the time-reversal symmetry can be broken by supercurrent induced by the phase differences and the spatial-inversion symmetry can be broken by tuning both the phase differences of two JJs asymmetric. The symmetry breaking of the short-range coupled JJs provides important insights for bottom-up engineering of novel SC phenomena in JJ arrays [24]. Experimental observation of SDE can elucidate the physics of symmetry breaking in coupled JJs and thus contribute to the engineering of novel SC phenomena. Furthermore, unlike the SDE studied thus far, the SDE in the two coupled JJs can be controlled non-locally by tuning the non-local phase difference. This means that the symmetry breaking can be realized only by phase engineering, which provides new designability and functionality for SC circuit applications.

There are several reports of nonreciprocal supercurrent transport or SDE measured with more than two current sources [25–30] or a single current source with the magnetic field [31] in multiterminal JJs which have more than three SC electrodes contacted to the same single normal metal. Compared to the multiterminal JJ structures in the literature, the coupled JJ structure utilizes the overlapping of the Andreev bound state wavefunctions through the SC electrode shared by two JJs (see Fig. 1(a)). Thanks to this structure, the coherent coupling effect on SDE is clearly distinguished and therefore provides the first evidence of coherent coupling to generate the SDE. The coherent coupling might also exist in the multiteminal JJ devices [31]. However, in these devices the coherent coupling contributions in the observed SDE cannot be distinguished from the SDE derived from the multiple JJs with no coherent coupling represented as the equivalent circuit in the literature [29,31], and no argument is carried out about the coherent coupling. Furthermore, the coupled JJ structure allows the coherent coupling of the JJs of the different normal metals. This property will lead to generating novel SC phenomena in combinations of various JJs which cannot be realized in the multiterminal JJ structure. Therefore, it is significant to study the fundamental physics of symmetry breaking in the coupled JJ by measuring the SDE.

We herein report on the observation of SDE in a JJ coupled to an adjacent JJ with a fixed phase difference, which is realized without Zeeman splitting or use of ferromagnetic materials. The SDE systematically appears in the vicinity of the $\pi$ non-local phase difference points. We observe the SDE with a single current source and phase engineering by magnetic fields. This is the difference from previous reports of the multiterminal JJs in which more than two current sources have been used (see supplementary note 9 and supplementary Figure S14). The coupled JJ structure enables to measure the SDE in the coupled structure of two different JJs realized by gate tuning. Thanks to this property of the structure, the local and non-local gate-voltage dependences elucidate that the SDE is enhanced when the two JJs become nearly equivalent. We attribute the origin of this SDE to the asymmetric current phase relation (CPR) resulting from short-range coherent coupling that breaks the time-reversal and spatial-inversion symmetries. Our results demonstrate a new mechanism for inducing the SDE that is applicable to dissipationless rectifiers in SC circuits and imply the coupled JJs as possible platforms to realize exotic SC phenomena realized under the time-reversal and spatial-inversion symmetries broken.

**Device characteristics**

*Physical Description*

In our experiments, a high-quality InAs quantum well covered by an epitaxial aluminum (Al) thin film has been used to form superconductor–semiconductor junctions [32]. This system provides a suitable platform for studying the SC proximity effect in the semiconductor owing to the highly transparent interface between the Al film and the InAs quantum well [33,34].

This quantum well wafer has been processed to fabricate a device with two coupled JJs (JJ1 and JJ2). To control the phase difference of JJ2 by an out-of-plane magnetic field, JJ1 and JJ2 are embedded in different SC loops with the LJJ1 and LJJ2 which are larger than JJ1 and JJ2, respectively. A schematic of the device is shown in Fig. 1(b). All JJs are gate-tunable using voltages to pinch off the junctions. JJ1 and JJ2 have the same structure (junction length and width are 100 nm and 600 nm, respectively), and the distance between them is 150 nm. This distance is much shorter than the coherence length of Al (approximately 1 μm). LJJ1 and LJJ2 are 2 μm wide and 100 nm in length (see Supplementary Note 1 and Figure S1 for additional details). For the results discussed in the main text, we have always pinched off LJJ1. Therefore, the device can be regarded as JJ1 with no loop and JJ2 embedded in the SC loop with LJJ2. Electron transport measurements have been performed at a base temperature of 10 mK using a dilution refrigerator. We have measured JJ1 by sweeping current $I_1$ to detect the voltage difference $V_1$, as shown in Fig. 1(b). This setup is completely different from that of the conventional switching current measurement for SQUIDs in which two JJs are embedded in the SC loop. In conventional SQUID measurements, the voltage difference of the two JJs is detected, whereas in our case, JJ1 is placed outside of the SC loop, and the bias current flows into the center SC electrode that JJ1 and JJ2 share. This means that the voltage differences of JJ2 and LJJ2 are always zero (see Supplementary Note 5 and Figures S7-S9).

*Supercurrent Analysis*

The gate voltage dependence is examined to verify that the supercurrent flows in the JJs through the InAs quantum well. Figure 1(c) shows $V_1$ as a function of $I_1$ and gate voltage $V_{g1}$ for the single JJ1 with JJ2, LJJ1, and LJJ2 pinched off. The supercurrent region where $V_1 \approx 0$ mV disappears at approximately $V_{g1} = -1.8$ V. Similarly, the gate voltage dependence for the current through JJ2 is depicted in Fig 1(d). Evidently, a negative $V_{g2}$ can suppress the supercurrent through JJ2. These results indicate that the gate voltages on the respective JJs can be used to control the supercurrent flow in the InAs quantum well.

We examine the oscillation of the switching current of JJ1 coupled to JJ2 embedded in the SC loop. In the device, the supercurrent of JJ1 depends not only on the phase difference of JJ1 but also on the phase difference of JJ2 due to the coherent coupling between the two JJs [20]. Therefore, the switching current is expected to oscillate as a function of the magnetic field $B$ because the phase difference of JJ2 is controlled by the magnetic flux in the loop [23]. We have measured this oscillation in our device with $V_{g1}$ = 0 V, $V_{g2}$ = 0 V, LJJ1 off, and LJJ2 on. LJJ2 has a sufficiently larger switching current than JJ2 and then $I_1$ in the range of the JJ1 switching current measurement hardly shifts the phase difference of LJJ2. (see Supplementary Note 4). Figure 2(a) shows $V_1$ as a function of $I_1$ and $B$. The white region ($V_1$ = 0 mV) indicates supercurrent flow in JJ1. The boundary between the white and red regions defines $I_{sw+}$, which is the switching current in the positive-current region. Similarly, $I_{sw-}$ (the switching current in the negative-current region) is defined as the boundary between the white and blue regions. Both $I_{sw+}$ and $I_{sw-}$ oscillate as a function of $B$, as expected from the coherent coupling between JJ1 and JJ2. We note that the coherence is necessary to obtain these oscillations and then the oscillation disappears in the device with 1 μm separation of JJ1 and JJ2 (see Supplementary Note 8 and Figure S13). The oscillation period is 0.156 mT, which is consistent with the calculated period of 0.176 mT for a loop area of 11.7 μm².

To verify the SDE, $V_1$ vs. $|I_1|$ at $B$ = 0.078 mT and 0.090 mT is plotted selectively by the purple and orange lines, respectively, in Fig. 2(b). The circles and squares represent the data obtained as $I_1$ is swept from zero to positive and negative values, respectively. At $B$ = 0.078 mT, $|I_{sw+}|$ = 106 nA (evaluated from the red circles) is smaller than $|I_{sw-}|$ = 116 nA (evaluated from the red squares). Conversely, at $B$ = 0.090 mT, $|I_{sw+}|$ = 117 nA> $|I_{sw-}|$ = 105 nA. These results confirm that the SDE occurs in JJ1. Therefore, the sign of the SDE, namely, the sign of $I_{sw+}$ + $I_{sw-}$, depends on $B$ (see Supplementary Note 2 and Figure S2).

To investigate the $B$-dependence of the SDE, the absolute values of $I_{sw+}$ and $I_{sw-}$ are plotted as a function of $B$ in Fig. 2(c), where the red and blue circles represent $I_{sw+}$ and $I_{sw-}$, respectively. For clarity, the regions where $|I_{sw+}|$ > $|I_{sw-}|$ are colored in red, and those where $|I_{sw+}|$ < $|I_{sw-}|$ are indicated in blue. The SDE ($|I_{sw+}|\neq |I_{sw-}|$) systematically appears in the vicinity of the minimal $|I_{sw}|$ points. This implies that the SDE sign can be easily switched by a change in $B$. Because of the time-reversal relation of the supercurrent in JJ1, $I_{sc1}$ of the supercurrent in JJ1 should satisfy $I_{sc1}(\phi_1, \phi_2) = -I_{sc1}(-\phi_1, -\phi_2)$, which results in $I_{sc1}(\phi_1, 0) = -I_{sc1}(-\phi_1, 0)$ and $I_{sc1}(\phi_1, \pi) = -I_{sc1}(-\phi_1, \pi)$. Here, $\phi_1$ and $\phi_2$ are the phase differences of JJ1 and JJ2, respectively. Therefore, we can set $\phi_2$ = 0 and $\pi$ as crossing points of $|I_{sw+}(B)|$ and $|I_{sw-}(B)|$, as indicated by the top axis in Fig. 2(c). It is significant that the condition $I_{sc1}(\phi_1, \phi_2) = -I_{sc1}(-\phi_1, \phi_2)$ is not necessarily satisfied when $\phi_2 \neq 0, \pi$ (mod $2\pi$), which corresponds to the broken time-reversal symmetry. In addition, the upper and central SC electrodes of JJ1 are not equivalent because the central SC electrode is coupled to the adjacent JJ, which induces a spatial-inversion symmetry break. Thus, the SDE can appear under the condition of $\phi_2 \neq 0, \pi$ (mod $2\pi$). Notably, a device with a mirrored structure produces an SDE with the opposite sign (see Supplementary Note 3 and Figure S3). The observed SDE features can be reproduced in a different device experimentally (see supplementary Note 7 and Figures S11 and S12) and also by numerical calculations based on the tight-binding model (see Supplementary Note 6 and Figure S10).

*Gate voltage effects on SDE*

Then, we study the gate voltage dependences of the SDE to verify that the SDE originates from coherent coupling and excludes other possible mechanisms. Figure 3(a) presents $|I_{sw+}|$ and $|I_{sw-}|$ as a function of $B$ at $V_{g2} = -1.4, -1.6$, and $-2.0$ V with $V_{g1} = 0$ V, LJJ1 off, and LJJ2 on, corresponding to the non-local gate control of JJ1 through the coherent coupling. As $V_{g2}$ becomes more negative, the oscillation amplitude and the SDE decrease. Finally, when the supercurrent in JJ2 disappears for $V_{g2} \leq -1.8$ V as indicated in Fig. 1(d), the SDE vanishes and the oscillation disappears, and only the supercurrent through JJ1 with no coupling remains. This demonstrates that the observed SDE is generated only when JJ2 is switched on and is therefore not attributed to any possible mechanisms that could induce SDE in single JJs [9] or to a possible vortex insertion in the SC electrodes [3,35]. The maximum $|I_{sw}|$ (in the $\phi_2 = 0$ (mod $2\pi$) neighbourhood; $B \sim 0.01, 0.168$ mT) in the coupled case ($V_{g2} = -1.4$ V and $-1.6$ V) exceeds $|I_{sw}|$ in the single JJ1 case ($V_{g2} = -2.0$ V). This is one of the expected features occurring as a consequence of the coherent coupling between JJ1 and JJ2. The coherent coupling of JJ1 and JJ2 is formed when the wavefunction penetration length is longer than the centre electrode width, as shown in Fig. 1(a). When JJ2 is switched off, the partial wave functions of JJ1 penetrating JJ2 disappear and do not contribute to the supercurrent in JJ1. Therefore, the switching current with JJ2 off is smaller than that when JJ2 is on at $\phi_2 = 0$.

Figure 3(b) shows the local gate control of $|I_{sw}|$ in JJ1 with respect to the oscillation and SDE at $V_{g1} = -1.2$ to $-1.6$ V with $V_{g2} = 0$ V, LJJ1 off, and LJJ2 on. As $V_{g1}$ becomes more negative, both $|I_{sw+}|$ and $|I_{sw-}|$ decrease. The results also indicate that a more negative $V_{g1}$ decreases the SDE remarkably. We define the SDE ratio as $\eta \equiv \frac{|I_{sw+}|-|I_{sw-}|}{|I_{sw+}|+|I_{sw-}|}$ and focus on $B = 0.0768$ mT, where the largest SDE for $|I_{sw+}| < |I_{sw-}|$ is obtained at $V_{g2} = -1.4$ V in Fig. 3(a) highlighted by the black arrow. In this case, $\eta = -0.064$ at $V_{g1} = -1.2$ V, whereas $\eta = -0.016$ at $V_{g1} = -1.6$ V. Notably, a negative $\eta$ is obtained for $|I_{sw+}| < |I_{sw-}|$ at $B = 0.0768$ mT; therefore, the larger SDE at $B = 0.0768$ mT is represented by the more negative $\eta$. This result reveals that the observed SDE does not originate from the asymmetric SQUID formed by JJ2 and LJJ2. This is because if this were the case [12,36], $\eta$ would not depend on the local gate voltage of JJ1, which is outside of the SQUID.

To clarify the relationship between the observed SDE and the correlation between JJ1 and JJ2, we fix $B = 0.0768$ mT and measure $I_{sw+}$ and $I_{sw-}$ as a function of $V_{g1}$ and $V_{g2}$. Typical results for $|I_{sw+}|$ and $|I_{sw-}|$ as a function of $V_{g2}$ at $V_{g1} = -1.25, -1.375$, and $-1.5$ V with LJJ1 off and LJJ2 on are shown in Fig. 3(c). At $V_{g1} = -1.25$ V and $-1.375$ V, the SDE decreases monotonically as $V_{g2}$ decreases. On the other hand, the SDE at $V_{g1}$ = −1.5 V becomes the largest with η = −0.075 at $V_{g2}$ = −1.525 V. Figure 3(d) shows the evaluated η as a function of $V_{g1}$ and $V_{g2}$ at $B$ = 0.0768 mT. The blue region corresponding to the large SDE appears diagonally on the $V_{g1}$ - $V_{g2}$ graph. This diagonal SDE dependence seen in Fig. 3(d) produces that η, namely the SDE at $V_{g1}$ = −1.5 V nonmonotonically depends on $V_{g2}$ in Fig. 3(c). We indicate the gate conditions ($V_{g1}$, $V_{g2}$) where $I_{sw}$ in the single JJ1 case (with JJ2 off) is equal to $I_{sw}$ in the single JJ2 case (with JJ1 off) as a grey curve in Fig. 3(d). Consequently, the blue SDE region clearly follows the grey curve, indicating that the symmetric condition of JJ1 and JJ2 is significant for observing the SDE in our device. The mirrored structure device also indicates a similar gate dependence of the SDE ratio with the sign reversed (see Supplementary Note 3 and Figure S4). The coherent coupling of the two JJs is formed when the Andreev bound states in JJ1 and JJ2 hold the same energies. Thus, when the number of bound states, energies, and transmissions of the JJs are similar or the same, the coherent coupling affects the supercurrent transport more remarkably. Therefore, the symmetric condition of the JJs' switching current that produces a larger SDE in the experiments is reasonable for the SDE originating from the coherent coupling of the JJs. Further, the inductance effect cannot generate such gate dependence of η because the SDE generated by the inductance and the bias current $I_1$ decreases monotonically as $I_1$ becomes smaller. Thus, if the SDE were produced by the inductance of the SC loop, only the local gate control would monotonously decrease the SDE; therefore, the diagonal feature shown in Fig. 3(d), would not be observed (see Supplementary Note 4 and Figure S5-6).

*SDE Origins*

Finally, we herein discuss the possible microscopic mechanism of the observed SDE derived from the coherent coupling. It has been theoretically proposed that elastic cotunneling (EC) and crossed Andreev reflection (CAR) coexist as a mechanism to form the short-range coherent coupling [20]. Here, cotunneling implies quasiparticle tunneling through the central SC electrode from JJ1 to JJ2, whereas the crossed Andreev reflection [37–39] describes the tunneling of an electron from JJ1 into JJ2 as a hole. Owing to the two different coupling energies, the Andreev bound states (as a function of $\phi_1$ with a fixed $\phi_2 \neq 0, \pi$) become asymmetric to $\phi_1 = 0$, resulting in the asymmetric CPR of JJ1 because the supercurrent is proportional to the differential of the energies created by $\phi_1$ [20]. This asymmetric CPR allows for the difference in positive and negative critical currents, thus providing the SDE.

In the literature [20], the predicted CPR indicates a finite supercurrent at $\phi_1 = 0$ (namely at the ϕ junction), in which the ground state of a JJ appears at a finite phase difference far from 0 or π [40–46]. Therefore, the obtained SDE can be an experimental signature of the ϕ junction formation in JJ1. The ϕ junction is expected to be applied to cryogenic memory cells [47] and phase batteries [48,49]. In addition, our results can be explained without considering the spin-orbit interactions present in InAs quantum wells. Recently such spin-orbit interactions have been reported to play an important role in the SDE observed in single JJs in the presence of an in-plane magnetic field [9]. Therefore, the in-plane magnetic field may be utilized to study the physics of spin-orbit interactions in the SDE in the coupled JJs. Further experimental studies to reveal the ϕ junction and the spin-orbit interaction roles in the SDE in coupled JJs are required to establish the coupled JJ physics and to evaluate any potential applications.

To establish these coupled JJ physics, it is demanded to observe the Andreev spectrum of the coupled JJs. This will be realized by the tunnel spectroscopy of the respective JJs as the previous report of the multiterminal JJs [50]

**Conclusion**

We demonstrate that the SDE in a JJ controlled by the non-local phase difference emerges from the coherent coupling of two JJs. The local and non-local gate voltage dependences indicate that the SDE is most notable when the two JJs become nearly equivalent, implying that the observed SDE is derived from the coherent coupling between two JJs. The SDE observations indicate that there are two different coupling mechanisms, EC and CAR, possessing different coupling energies. Our results indicate that the time reversal and spatial inversion symmetries in one JJ are broken by non-local control of the other coupled JJ. This means that the coupled JJs will be platforms to exploit the exotic SC phenomena which have been demonstrated with the materials of spin-orbit interactions and the strong magnetic fields and enable the design of novel SC phenomena with potential applications in SC diode devices.

# References

[1] R. M. Lutchyn, E. P. A. M. Bakkers, L. P. Kouwenhoven, P. Krogstrup, C. M. Marcus, and Y. Oreg, *Majorana Zero Modes in Superconductor–Semiconductor Heterostructures*, Nature Reviews Materials **3**, 52 (2018).

[2] F. Raissi and J. E. Nordman, *Josephson Fluxonic Diode*, Applied Physics Letters **65**, 1838 (1994).

[3] G. Carapella and G. Costabile, *Ratchet Effect: Demonstration of a Relativistic Fluxon Diode*, Physical Review Letters **87**, 077002 (2001).

[4] M. Beck, E. Goldobin, M. Neuhaus, M. Siegel, R. Kleiner, and D. Koelle, *High-Efficiency Deterministic Josephson Vortex Ratchet*, Physical Review Letters **95**, 090603 (2005).

[5] A. Sterck, R. Kleiner, and D. Koelle, *Three-Junction SQUID Rocking Ratchet*, Physical Review Letters **95**, 177006 (2005).

[6] A. Sterck, D. Koelle, and R. Kleiner, *Rectification in a Stochastically Driven Three-Junction SQUID Rocking Ratchet*, Physical Review Letters **103**, 047001 (2009).

[7] H. Sickinger, A. Lipman, M. Weides, R. G. Mints, H. Kohlstedt, D. Koelle, R. Kleiner, and E. Goldobin, *Experimental Evidence of A $\upvarphi$ Josephson Junction*, Physical Review Letters **109**, 107002 (2012).

[8] R. Menditto, H. Sickinger, M. Weides, H. Kohlstedt, D. Koelle, R. Kleiner, and E. Goldobin, *Tunable $\upvarphi$ Josephson Junction Ratchet*, Physical Review E **94**, 042202 (2016).

[9] C. Baumgartner et al., *Supercurrent Rectification and Magnetochiral Effects in Symmetric Josephson Junctions*, Nat. Nanotechnol. **17**, 39 (2022).

[10] B. Turini, S. Salimian, M. Carrega, A. Iorio, E. Strambini, F. Giazotto, V. Zannier, L. Sorba, and S. Heun, *Josephson Diode Effect in High Mobility InSb Nanoflags*, Nano Lett. **22**, 8502 (2022).

[11] B. Pal et al., *Josephson Diode Effect from Cooper Pair Momentum in a Topological Semimetal*, Nat. Phys. **18**, 1228 (2022).

[12] M. D. Thompson, M. B. Shalom, A. K. Geim, A. J. Matthews, J. White, Z. Melhem, Y. A. Pashkin, R. P. Haley, and J. R. Prance, *Graphene-Based Tunable SQUIDs*, Applied Physics Letters **110**, 162602 (2017).

[13] A. Murphy and A. Bezryadin, *Asymmetric Nanowire SQUID: Linear Current-Phase Relation, Stochastic Switching, and Symmetries*, Physical Review B **96**, 094507 (2017).

[14] R. Wakatsuki, Y. Saito, S. Hoshino, Y. M. Itahashi, T. Ideue, M. Ezawa, Y. Iwasa, and N. Nagaosa, *Nonreciprocal Charge Transport in Noncentrosymmetric Superconductors*, Science Advances **3**, e1602390 (2017).

[15] F. Qin, W. Shi, T. Ideue, M. Yoshida, A. Zak, R. Tenne, T. Kikitsu, D. Inoue, D. Hashizume, and Y. Iwasa, *Superconductivity in a Chiral Nanotube*, Nature Communications **8**, 14465 (2017).

[16] F. Ando, Y. Miyasaka, T. Li, J. Ishizuka, T. Arakawa, Y. Shiota, T. Moriyama, Y. Yanase, and T. Ono, *Observation of Superconducting Diode Effect*, Nature **584**, 373 (2020).

[17] Y. Miyasaka et al., *Observation of Nonreciprocal Superconducting Critical Field*, Applied Physics Express **14**, 073003 (2021).

[18] J. Diez-Merida, A. Diez-Carlon, S. Y. Yang, Y.-M. Xie, X.-J. Gao, K. Watanabe, T. Taniguchi, X. Lu, K. T. Law, and D. K. Efetov, *Magnetic Josephson Junctions and Superconducting Diodes in Magic Angle Twisted Bilayer Graphene*, arXiv:2110.01067.

[19] J.-X. Lin, P. Siriviboon, H. D. Scammell, S. Liu, D. Rhodes, K. Watanabe, T. Taniguchi, J. Hone, M. S. Scheurer, and J. I. A. Li, *Zero-Field Superconducting Diode Effect in Small-Twist-Angle Trilayer Graphene*, Nat. Phys. **18**, 1221 (2022).

[20] J.-D. Pillet, V. Benzoni, J. Griesmar, J.-L. Smirr, and Ç. O. Girit, *Nonlocal Josephson Effect in Andreev Molecules*, Nano Letters **19**, 7138 (2019).

[21] V. Kornich, H. S. Barakov, and Y. V. Nazarov, *Fine Energy Splitting of Overlapping Andreev Bound States in Multiterminal Superconducting Nanostructures*, Physical Review Research **1**, 033004 (2019).

[22] V. Kornich, H. S. Barakov, and Y. V. Nazarov, *Overlapping Andreev States in Semiconducting Nanowires: Competition of One-Dimensional and Three-Dimensional Propagation*, Physical Review B **101**, 195430 (2020).

[23] S. Matsuo, J. S. Lee, C.-Y. Chang, Y. Sato, K. Ueda, C. J. Palmstrøm, and S. Tarucha, *Observation of Nonlocal Josephson Effect on Double InAs Nanowires*, Commun Phys **5**, 221 (2022).

[24] J. D. Sau and S. D. Sarma, *Realizing a Robust Practical Majorana Chain in a Quantum-Dot-Superconductor Linear Array*, Nature Communications **3**, 1966 (2012).

[25] A. W. Draelos, M.-T. Wei, A. Seredinski, H. Li, Y. Mehta, K. Watanabe, T. Taniguchi, I. V. Borzenets, F. Amet, and G. Finkelstein, *Supercurrent Flow in Multiterminal Graphene Josephson Junctions*, Nano Letters **19**, 1039 (2019).

[26] N. Pankratova, H. Lee, R. Kuzmin, K. Wickramasinghe, W. Mayer, J. Yuan, M. G. Vavilov, J. Shabani, and V. E. Manucharyan, *Multiterminal Josephson Effect*, Physical Review X **10**, (2020).

[27] G. V. Graziano, J. S. Lee, M. Pendharkar, C. J. Palmstrøm, and V. S. Pribiag, *Transport Studies in a Gate-Tunable Three-Terminal Josephson Junction*, Physical Review B **101**, (2020).

[28] E. G. Arnault, S. Idris, A. McConnell, L. Zhao, T. F. Q. Larson, K. Watanabe, T. Taniguchi, G. Finkelstein, and F. Amet, *Dynamical Stabilization of Multiplet Supercurrents in Multiterminal Josephson Junctions*, Nano Lett. **22**, 7073 (2022).

[29] J. Chiles, E. G. Arnault, C.-C. Chen, T. F. Q. Larson, L. Zhao, K. Watanabe, T. Taniguchi, F. Amet, and G. Finkelstein, *Non-Reciprocal Supercurrents in a Field-Free Graphene Josephson Triode*, arXiv:2210.02644.

[30] F. Zhang, A. S. Rashid, M. T. Ahari, W. Zhang, K. M. Ananthanarayanan, R. Xiao, G. J. de Coster, M. J. Gilbert, N. Samarth, and M. Kayyalha, *Andreev Processes in Mesoscopic Multi-Terminal Graphene Josephson Junctions*, arXiv:2210.04408.

[31] M. Gupta, G. V. Graziano, M. Pendharkar, J. T. Dong, C. P. Dempsey, C. Palmstrøm, and V. S. Pribiag, *Superconducting Diode Effect in a Three-Terminal Josephson Device*, arXiv:2206.08471.

[32] F. Nichele et al., *Scaling of Majorana Zero-Bias Conductance Peaks*, Physical Review Letters

**119**, 136803 (2017).

[33] M. Kjaergaard et al., *Quantized Conductance Doubling and Hard Gap in a Two-Dimensional Semiconductor?Superconductor Heterostructure*, Nature Communications **7**, 12841 (2016).

[34] M. Kjaergaard, H. J. Suominen, M. P. Nowak, A. R. Akhmerov, J. Shabani, C. J. Palmstrøm, F. Nichele, and C. M. Marcus, *Transparent Semiconductor-Superconductor Interface and Induced Gap in an Epitaxial Heterostructure Josephson Junction*, Phys. Rev. Applied **7**, 034029 (2017).

[35] T. Golod and V. M. Krasnov, *Demonstration of a Superconducting Diode-with-Memory, Operational at Zero Magnetic Field with Switchable Nonreciprocity*, Nat Commun **13**, 3658 (2022).

[36] R. S. Souto, M. Leijnse, and C. Schrade, *The Josephson Diode Effect in Supercurrent Interferometers*, arXiv:2205.04469.

[37] S. G. den Hartog, C. M. A. Kapteyn, B. J. van Wees, T. M. Klapwijk, and G. Borghs, *Transport in MultiTerminal Normal-Superconductor Devices: Reciprocity Relations, Negative and Nonlocal Resistances, and Reentrance of the Proximity Effect*, Physical Review Letters **77**, 4954 (1996).

[38] R. Mélin and D. Feinberg, *Sign of the Crossed Conductances at a Ferromagnet/Superconductor/Ferromagnet Double Interface*, Physical Review B **70**, 174509 (2004).

[39] S. Russo, M. Kroug, T. M. Klapwijk, and A. F. Morpurgo, *Experimental Observation of Bias-Dependent Nonlocal Andreev Reflection*, Physical Review Letters **95**, 027002 (2005).

[40] A. Buzdin and A. E. Koshelev, *Periodic Alternating 0- And $\uppi$-Junction Structures as Realization Of $\upvarphi$-Josephson Junctions*, Physical Review B **67**, 220504 (2003).

[41] A. Buzdin, *Direct Coupling Between Magnetism and Superconducting Current in the Josephson $\upvarphi 0$ Junction*, Physical Review Letters **101**, 107005 (2008).

[42] Y. Tanaka, T. Yokoyama, and N. Nagaosa, *Manipulation of the Majorana Fermion, Andreev Reflection, and Josephson Current on Topological Insulators*, Physical Review Letters **103**, 107002 (2009).

[43] A. A. Reynoso, G. Usaj, C. A. Balseiro, D. Feinberg, and M. Avignon, *Anomalous Josephson Current in Junctions with Spin Polarizing Quantum Point Contacts*, Physical Review Letters **101**, 107001 (2008).

[44] T. Yokoyama, M. Eto, and Y. V. Nazarov, *Anomalous Josephson Effect Induced by Spin-Orbit Interaction and Zeeman Effect in Semiconductor Nanowires*, Physical Review B **89**, 195407 (2014).

[45] D. B. Szombati, S. Nadj-Perge, D. Car, S. R. Plissard, E. P. A. M. Bakkers, and L. P. Kouwenhoven, *Josephson $\upphi 0$-Junction in Nanowire Quantum Dots*, Nature Physics **12**, 568 (2016).

[46] W. Mayer, M. C. Dartiailh, J. Yuan, K. S. Wickramasinghe, E. Rossi, and J. Shabani, *Gate Controlled Anomalous Phase Shift in Al/InAs Josephson Junctions*, Nature Communications **11**,

212 (2020).

[47] E. Goldobin, H. Sickinger, M. Weides, N. Ruppelt, H. Kohlstedt, R. Kleiner, and D. Koelle, *Memory Cell Based on a $\upvarphi$ Josephson Junction*, Applied Physics Letters **102**, 242602 (2013).

[48] S. Pal and C. Benjamin, *Quantized Josephson Phase Battery*, EPL (Europhysics Letters) **126**, 57002 (2019).

[49] E. Strambini et al., *A Josephson Phase Battery*, Nature Nanotechnology **15**, 656 (2020).

[50] E. Strambini, S. D'Ambrosio, F. Vischi, F. S. Bergeret, Yu. V. Nazarov, and F. Giazotto, *The ω-SQUIPT as a Tool to Phase-Engineer Josephson Topological Materials*, Nature Nanotech **11**, 1055 (2016).

**Figures**

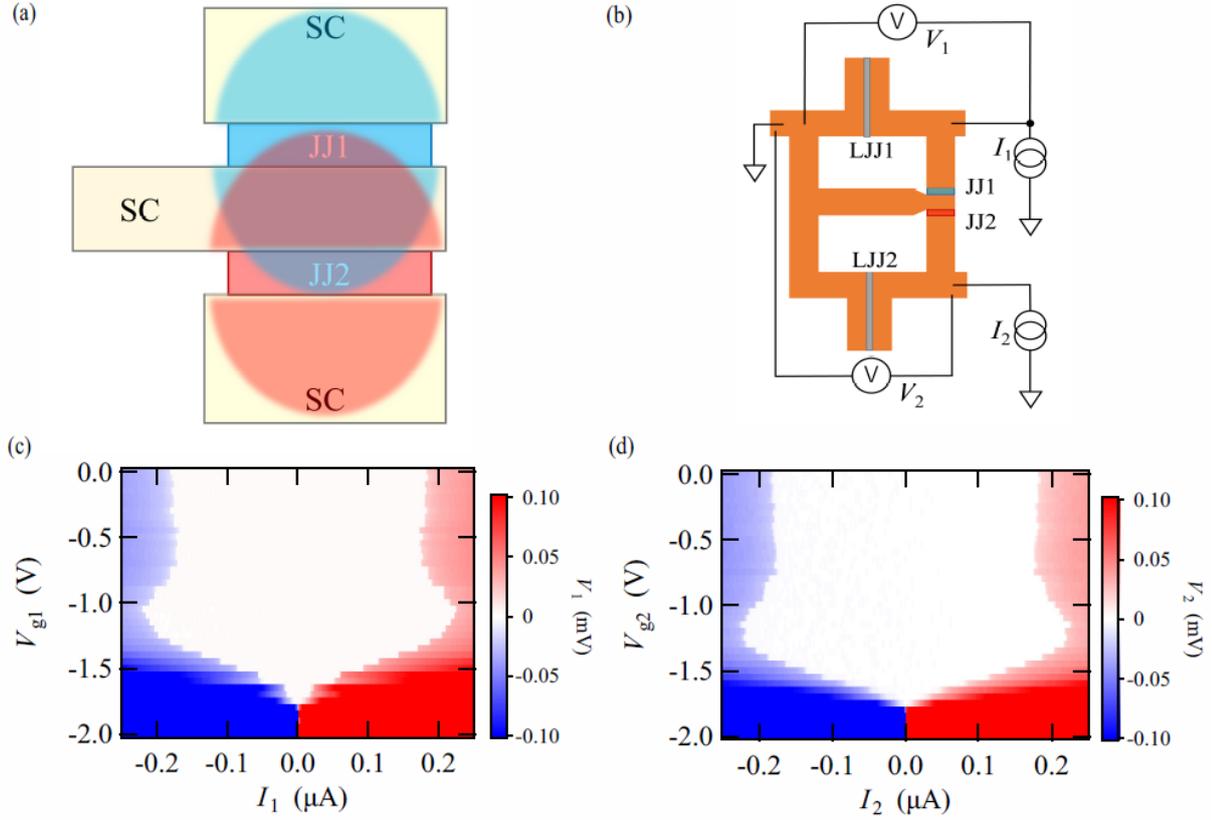

**Fig. 1: Device concept and coherent coupling of JJs**

(a) Conceptual image of the coherent coupling of two JJs (JJ1 and JJ2). The wave functions of the Andreev bound states in the respective JJs penetrate the SC electrodes. When the center electrode is sufficiently thin to allow the wave functions to tunnel into the adjacent JJ, the bound states in the JJs hybridize, resulting in coherent coupling.

(b) Schematic image of our device and measurement setup. Two JJs (JJ1 and JJ2) are fabricated from an InAs quantum well covered by an epitaxial Al film. Additionally, two larger JJs (LJJ1 and LJJ2) are fabricated. When acquiring the data, LJJ1 is always pinched off. Then only JJ2 is embedded in the SC loop with LJJ2, which enables control of the phase difference of JJ2 by an out-of-plane magnetic field. The voltage difference $V_1$ on JJ1 is measured based on the bias current $I_1$.

(c) $V_1$ on JJ1 with JJ2, LJJ1, and LJJ2 pinched off as a function of $I_1$ and $V_{g1}$ when $B = 0$ mT. The supercurrent disappears at a gate voltage of approximately −1.8 V.

(d) Voltage difference $V_2$ on JJ2 with JJ1, LJJ1, and LJJ2 pinched off as a function of the bias current $I_2$ and $V_{g2}$ when $B = 0$ mT.

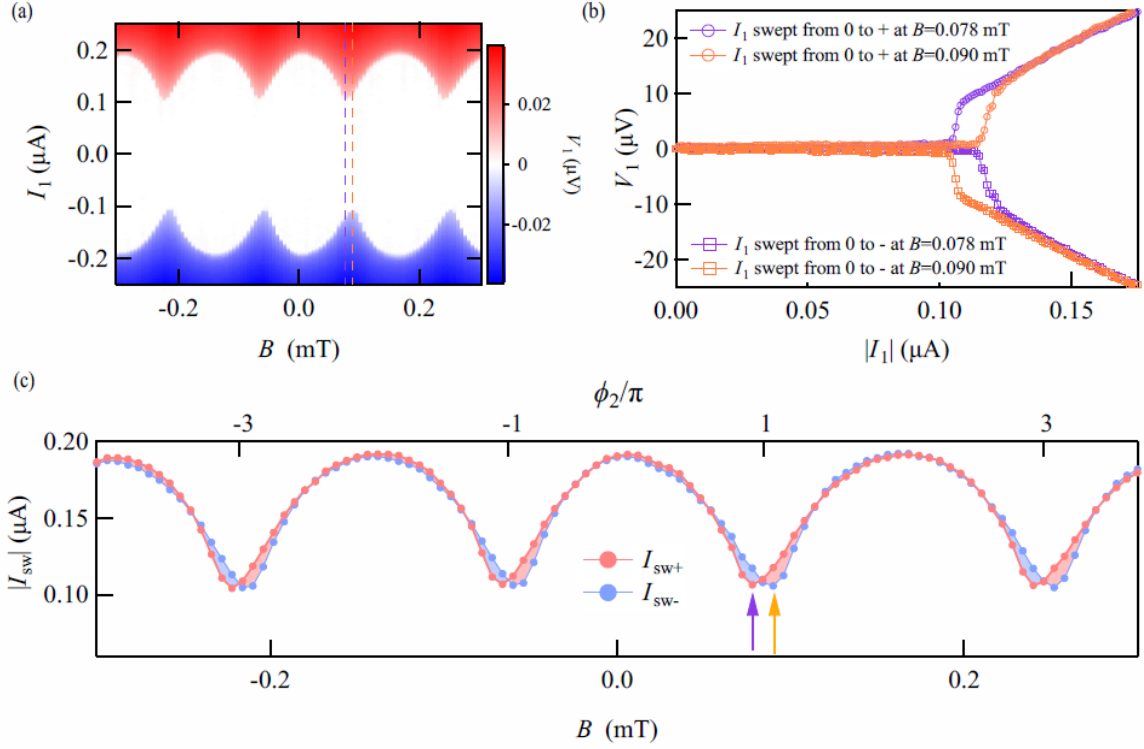

**Fig. 2: Observed superconducting diode effect (SDE)**
(a) $V_1$ as a function of $I_1$ and $B$ with $V_{g1} = 0$ V, $V_{g2} = 0$ V, LJJ1 off, and LJJ2 on. The oscillation derived from the coherent coupling of JJ1 and JJ2 is observed. The purple and orange dashed lines indicate results for $B = 0.078$ and $0.090$ mT, respectively.
(b) $V_1$ as a function of $|I_1|$ at $B = 0.078$ and $0.090$ mT are indicated by the purple and orange curves, respectively. The circles and squares are obtained as $I_1$ is swept from 0 to the positive and negative current directions, respectively. The results show that the switching current in the positive direction is different from that in the negative direction, indicating the presence of the SDE.
(c) $|I_{sw+}|$ and $|I_{sw-}|$ with respect to $B$ are represented by the red and blue circles, respectively, indicating that the SDE occurs systematically. In addition, the sign of the SDE is reversed (the blue and red shadows highlight the regions when $|I_{sw+}| < |I_{sw-}|$ and $|I_{sw+}| > |I_{sw-}|$, respectively). The purple and orange arrows indicate $B = 0.078$ and $0.090$ mT.

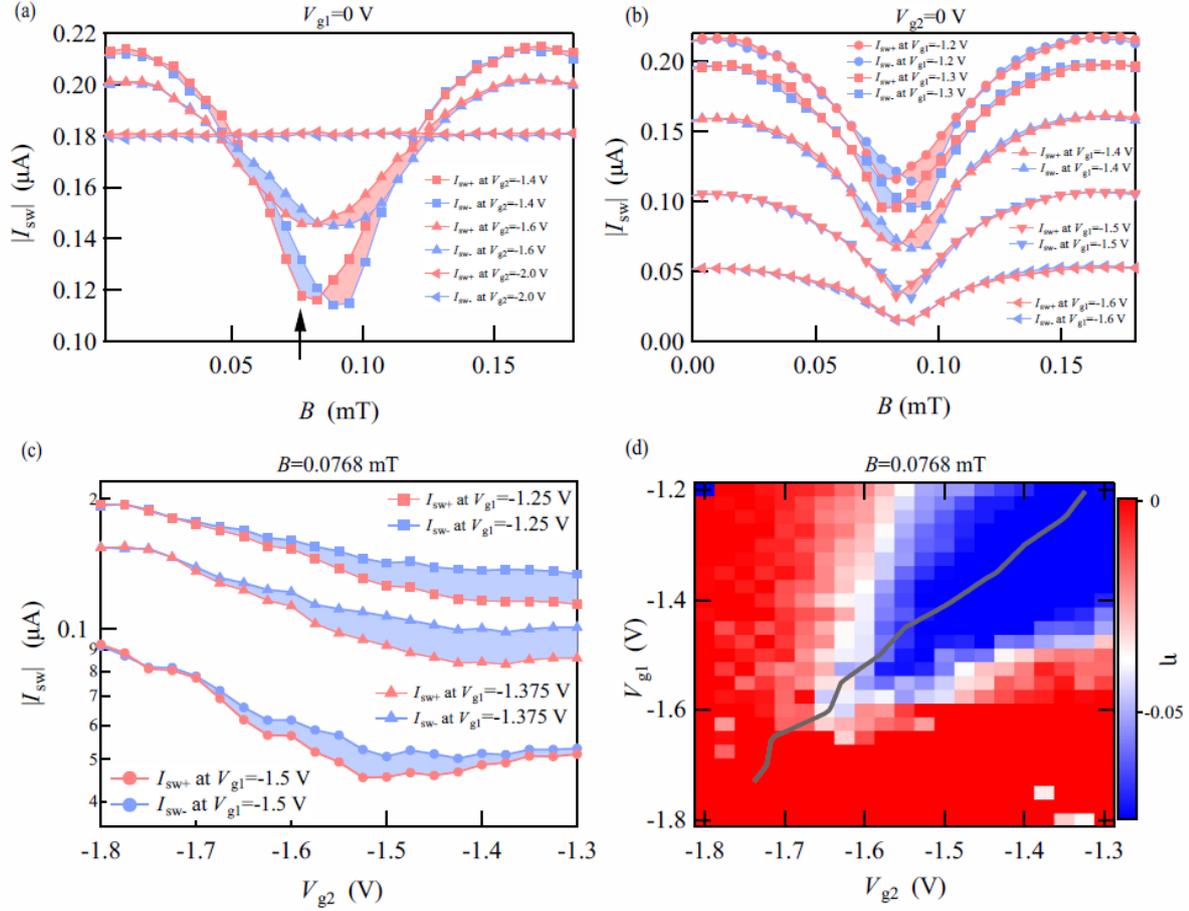

**Fig. 3: Symmetric conditions of JJ1 and JJ2 that generate the SDE**

(a) $|I_{sw+}|$ and $|I_{sw-}|$ as a function of $B$ at $V_{g2} = -1.4, -1.6$, and $-2.0$ V with $V_{g1} = 0$ V, LJJ1 off, and LJJ2 on. The SDE of JJ1 is non-locally controlled by tuning JJ2 with $V_{g2}$, and disappears as does the oscillation for $V_{g2} \leq -1.8$ V because the supercurrent in JJ2 disappears.

(b) $|I_{sw+}|$ and $|I_{sw-}|$ as a function of $B$ for $V_{g1} = -1.2$ to $-1.6$ V with $V_{g2} = 0$ V, LJJ1 off, and LJJ2 on. When JJ1 is locally controlled with $V_{g1}$, the SDE decreases as $V_{g1}$ decreases.

(c) $|I_{sw+}|$ and $|I_{sw-}|$ as a function of $V_{g2}$ at $V_{g1} = -1.25, -1.375$, and $-1.5$ V with LJJ1 off and LJJ2 on when $B = 0.0768$ mT. At $V_{g1} = -1.25$ and $-1.375$ V, the SDE monotonically decreases as $V_{g2}$ decreases. On the other hand, at $V_{g1} = -1.5$ V, the SDE becomes maximum at $V_{g2} = -1.525$ V.

(d) The evaluated SDE ratio η as a function of $V_{g1}$ and $V_{g2}$ at $B = 0.0768$ mT. The blue region represents when the SDE is enhanced, which is shaped diagonally. The grey curve indicates the gate condition when JJ1 and JJ2 have the same switching current. The blue SDE region follows the grey curve; this implies that the symmetric condition of JJ1 and JJ2 favors a larger SDE.

## Methods

*Sample growth*

The wafer structure has been grown via molecular beam epitaxy on a semi-insulating InP substrate. The stack materials from bottom to top are a 100 nm $In_{0.52}Al_{0.48}As$ buffer, a 5 period 2.5 nm $In_{0.53}Ga_{0.47}As$/2.5 nm $In_{0.52}Al_{0.48}As$ superlattice, a 1 μm thick metamorphic graded buffer stepped from $In_{0.52}Al_{0.48}As$ to $In_{0.84}Al_{0.16}As$, a 33 nm graded $In_{0.84}Al_{0.16}As$ to $In_{0.81}Al_{0.19}As$ layer, a 25 nm $In_{0.81}Al_{0.19}As$ layer, a 4 nm $In_{0.81}Ga_{0.19}As$ lower barrier, a 5 nm InAs quantum well, a 10 nm $In_{0.81}Ga_{0.19}As$ top barrier, two monolayers of GaAs, and finally an 8.7 nm layer of epitaxial Al. The top Al layer has been grown in the same chamber without breaking the vacuum. The two-dimensional electron gas (2DEG) accumulates in the InAs quantum well.

*Device Fabrication*

The JJs have been fabricated using conventional electron beam lithography with polymethyl methacrylate (PMMA). The aluminum film has been etched out using a type-D etchant after defining the mesa of the InAs quantum well with a 1:1:8 $H_3PO_4$, $H_2O_2$, and $H_2O$ etchant ratio. Then, a 30-nm-thick $Al_2O_3$ film has been grown by atomic layer deposition and Ti and Au have been deposited to obtain the gate electrodes.

## Acknowledgments

This work was partially supported by a JSPS Grant-in-Aid for Scientific Research (S) (Grant No. JP19H05610), JST PRESTO (grant no. JPMJPR18L8), JSPS Grant-in-Aid for Early-Career Scientists (Grant No. 18K13484), Advanced Technology Institute Research Grants, and the Ozawa-Yoshikawa Memorial Electronics Research Foundation.

## Author contributions

S.M. designed the experiments. T.L., S.G., G.C.G, and M.J.M grew wafers to form InAs 2DEG quantum wells covered with epitaxial aluminum. S.M. fabricated the devices. S.M. and T.I. performed measurements. S.M., T.I., Y.S., and S.T. analyzed the data. T.Y. performed numerical calculations. S.T. supervised the study.

## Competing interest statement

The authors declare no competing interests.

## Supplementary information

Supplementary Notes 1–9 and Supplementary Figures S1–14.

## Corresponding authors

Correspondence and requests should be addressed to Sadashige Matsuo (sadashige.matsuo@riken.jp) and Seigo Tarucha (tarucha@riken.jp).

# Supplementary Information

## Supplementary Note 1: Full device information

Figure S1(a) shows an illustration of the device. The left shows a schematic, and the right provides an optical image of a dummy device having the same structure as the real device. Two superconducting quantum interference devices (SQUIDs) are present. JJ1 and a larger JJ (LJJ1) form an asymmetric SQUID, and JJ2 and LJJ2 also form an asymmetric SQUID. LJJ1 and LJJ2 have the same structure. The two superconducting (SC) loops have the same loop area. For the measurements discussed in the main manuscript, we always pinch off LJJ1. For the single JJ2 measurements presented in Fig. 1(d), we pinch off JJ1, LJJ1, and LJJ2, vary current $I_2$, and measure $V_2$.

Figures S1(b) and (c) show the single LJJ1 and LJJ2 measurement results with the other JJs pinched off, respectively. The switching currents in LJJ1 and LJJ2 are approximately 800 nA and 900 nA with no gate voltages applied, respectively. When the gate voltages decrease, the JJs are pinched off after the switching current is enhanced once. The same effect can be observed in JJ1 and JJ2 in Figs. 1(c) and (d). We attribute this feature to mobility enhancement. When the carrier density decreases, the mobility reaches a maximum and then decreases, as reported in the literature [1].

Figures S1(d), (e), (f), and (g) show the magnetic-field dependence of the respective single JJs. The switching current is less dependent on the magnetic field $B$ because our measurement range for $B$ is much lower than the 0th lobe of the Fraunhofer pattern (34 mT for JJ1 and JJ2, and 10 mT for LJJ1 and LJJ2 can be calculated from the junction area).

## Supplementary Note 2: I-V curves and SDE in JJ1 measurements

Figure S2(a) shows the I-V curves observed for JJ1 with JJ2 and LJJ2 on and LJJ1 pinched off. The red and blue results are obtained at $B$ = 0.078 and 0.090 mT, respectively. The line indicates the results in $I_1$ swept from the negative to the positive current direction, whereas the squares are obtained when $I_1$ is swept from the positive to the negative current direction. No hysteresis is observed for the up and down sweeps of $I_1$ when comparing lines and squares with the same colors. This implies that JJ1 is in the overdamped regime.

Figure S2(b) shows the nonreciprocal response in JJ1 at $B$ = 0.078 mT. Notably, $B$ = 0.078 mT corresponds to $0 < \phi_2 < \pi$. For the detection, we bias $I_1$ = −112.5 nA for 9.1 s and then $I_1$ = 112.5 nA for 9.1 s ten times. The results clearly represent the nonreciprocal response of $V_1$, namely $V_1 \simeq 0$ mV at $I_1$ = −112.5 nA, whereas $V_1 > 0$ mV at $I_1$ = 112.5 nA. This nonreciprocal response can be switched by controlling the nonlocal phase difference $\phi_2$ to change $B$. Figure S2(c) shows the $V_1$ response at $B$ = 0.096 mT, which corresponds to $-\pi < \phi_2 < 0$. We bias $I_1$ = −116 nA for 4.4 s and then $I_1$ = 116 nA for 4.4 s ten times. In this case, $V_1 < 0$ mV at $I_1$ = −116 nA, whereas $V_1 \simeq 0$ mV at $I_1$ = 116 nA. The observed nonreciprocal response is reversed compared to that at $B$ = 0.078 mT.

**Supplementary Note 3: SDE observed in the mirror-inverted structure**

We investigate the SDE in the mirror-symmetry structure by measuring JJ2 with JJ1 and LJJ1 on and LJJ2 pinched off. Notably, the two SC loops have the same loop area. Measurements of the SDE are performed on JJ2 by applying a bias current $I_2$ to detect $V_2$. The obtained $V_2$ values are shown in Fig. S3(a) as a function of $I_2$ and $B$. The white region corresponds to the supercurrent region where $V_2$ vanishes. Similar to the results shown in Fig. 2(a), a switching current oscillation with respect to $B$ is observed. The absolute values of the determined switching currents $|I_{sw+}|$ and $|I_{sw-}|$ are indicated by the red and blue circles in Fig. S3(b), respectively. The SDE is reproduced in this mirrored structure.

Contrary to the results shown in Fig. 2(c), in this mirrored structure, $|I_{sw+}| > |I_{sw-}|$ is obtained for $0 < \phi_1 < \pi$, whereas $|I_{sw+}| < |I_{sw-}|$ is obtained for $-\pi < \phi_1 < 0$. Thus, the sign of η is opposite to that shown in Fig. 2(c). Owing to the mirror inversion of the structure, the sign of the nonlocal phase difference induced by the magnetic field is also inverted compared to that shown in Fig. 2(c). For instance, when the phase of the central electrode is assumed to be 0, the lower electrode has a positive phase for positive $B$, while the upper electrode has a negative phase, as shown in Fig. S1(a).

We also confirm that the SDE follows the symmetric condition of the switching currents in this mirrored structure. After we changed the cables for the measurement of the mirrored structure after the measurements used to obtain Figs. 2 and 3 in the main paper, the gate dependence of our device changed slightly. Then, we measure the switching current vs. the gate voltage in the single JJ1 and JJ2 cases, as shown in Figs. S4(a) and (b), respectively. The switching current disappears for gate voltages of approximately −1.9 V, which are slightly more negative than those in Figs. 1(c) and (d). Then, we fix $B = 0.0768$ mT and measure the SDE as a function of $V_{g1}$ and $V_{g2}$. The η values calculated from the results are shown in Fig. S4.(c). Similar to Fig. 3(d), the red SDE region appears diagonally and follows the grey curve that indicates the gate conditions when $I_{sw}$ in JJ1 = $I_{sw}$ in JJ2 evaluated from Figs. S4(a) and (b). The obtained η is positive, whereas η in Fig. 3(d) is negative. This also indicates that the SDE sign in the mirrored structure is reversed. These results further verify that the SDE appears when JJ1 and JJ2 are symmetric and support that the observed SDE is derived from the coherent coupling between JJ1 and JJ2.

**Supplementary Note 4: Evaluation of the inductance effect**

When current flows in the SC wire, the inductances can cause a change of magnetic flux in the SC loop. This can shift the $I_{sw}$ oscillation along the $B$ axis, and the shift direction becomes opposite when the bias current direction is reversed. Consequently, this inductance effect can also produce the SDE. In our results of Fig. 2(c), it is clarified that this inductance effect is not dominant in the observed SDE because the diagonal SDE signal in Fig. 3(d) cannot be explained by the SDE generated by the inductance effect. In order to further support this conclusion, we evaluate the SDE generated by the inductance effect and show that the contribution from the inductance to the observed SED is negligible.

First, the inductance effect in our SC loop is slightly different from that in conventional SQUID

measurements [2,3] because we measure JJ1 with JJ2 and LJJ2 embedded in the loop and LJJ1 pinched off in the main manuscript. In this setup, the bias current of $I_1$ to measure the switching current of JJ1 is distributed into the SC arm 1 and SC arm 2 as shown in Fig. S5(a) where the region below the center SC electrode between JJ1 and JJ2 in Fig. 1(b) is exhibited. The SC arm 1 is directly connected to the GND. Then the bias current does not necessarily flow through the JJs. Only the distributed current through the SC arm 2 makes the phase shifts of JJ2 and LJJ2. This situation is completely different from the setup for conventional SQUID switching current measurements as shown in Fig. S5(b). In the case shown in Fig. S5(b), the bias current distributes into two SC arms both of which include JJs (JJ2 and LJJ2).

We note that for the measurement of the JJ1 switching current to obtain the SDE such as in Fig. 2(c), we apply $I_1$ in the range of 200 nA determined by the JJ1 switching current. This is much smaller than that of around 1 μA, switching current of the SQUID of JJ2 and LJJ2.

To estimate the SDE generated by the inductance and remove it from the results in Fig. 2(c), we need to estimate how much phase shift of JJ2 can be generated in the setup of Fig. S5(a). Then we consider the equation of the phase differences of JJ2 and LJJ2. With no bias current, $\phi_{20} + \phi_{L20} = 2\pi(\Phi_{ext} - (L_{arm1} + L_{arm2})I_{cw})/\Phi_0$ is satisfied. Here $\phi_{20}, \phi_{L20}, \Phi_{ext}, L_{arm1}, L_{arm2},$ and $I_{cw}$ represent the phase difference of JJ2 with $I_1 = 0$, the phase difference of LJJ2 with $I_1 = 0$, the flux in the SC loop induced by the external magnetic field $B$, inductance of the SC arm 1, inductance of the SC arm 2, and the supercurrent in the SC loop dependent on the $\phi_{20}$ and $\phi_{L20}$, respectively. The left term represents the accumulated phases in JJ2 and LJJ2. The right term corresponds to the magnetic flux in the loop generated by the external magnetic fields and the self-inductance with the supercurrent in the loop.

When we apply $I_1$ to study the SDE, $I_1$ distributes into the two SC arms. We define $I_1 - \Delta I_{cw}$ as the distributed bias current flowing in the SC arm 1 to GND and $\Delta I_{cw}$ flowing in the SC arm 2, namely, JJ2 and LJJ2 to GND. Due to $\Delta I_{cw}$, the phase differences of JJ2 and LJJ2 change by $\Delta\phi_2$ and $\Delta\phi_{L2}$. For finite $I_1$, $(\phi_{20} + \Delta\phi_2) + (\phi_{L20} + \Delta\phi_{L2}) = 2\pi(\Phi_{ext} - L_{arm1}(I_{cw} + \Delta I_{cw} - I_1) - L_{arm2}(I_{cw} + \Delta I_{cw}))/\Phi_0$ is satisfied. Now we set the positive of the currents and phase differences as the arrow directions in Fig. S5(a).

Therefore, we can obtain:

$$\Delta\phi_2 + \Delta\phi_{L2} = \frac{2\pi(L_{arm1}I_1 - (L_{arm1} + L_{arm2})\Delta I_{cw})}{\Phi_0}. \qquad \text{Eq. 1}$$

The critical current of LJJ2 is much larger than that of JJ2, resulting in $|\Delta\phi_2| > |\Delta\phi_{L2}|$ because these phase shifts are induced by the same current of $\Delta I_{cw}$. In addition, when the current phase relation of JJ2 is written as $I_{JJ2}(\phi)$, $I_{cw} + \Delta I_{cw} = I_{JJ2}(\phi_{20} + \Delta\phi_2)$ with $I_{cw} = I_{JJ2}(\phi_{20})$ is obtained. Then, $\Delta I_{cw} \simeq \frac{dI_{JJ2}}{d\phi}(\phi_{20})\Delta\phi_2$. Using this and Eq. 1 with no $\Delta\phi_{L2}$, we can obtain

$$\Delta\phi_2 = \frac{L_{\text{arm1}} I_1}{\frac{\Phi_0}{2\pi} + (L_{\text{arm1}} + L_{\text{arm2}})\frac{dI_{\text{JJ2}}}{d\phi}(\phi_{20})} \qquad \text{Eq. 2}$$

Now our JJs are short ballistic JJs, then the current phase relation can be written as $I_{\text{JJ2}}(\phi) = \frac{A\tau \sin\phi}{\sqrt{1-\tau(\sin\frac{\phi}{2})^2}}$, where $A$ and $\tau$ represent the coefficient and transmission of the junction, respectively [3].

From the literature [3], $\tau$ evaluated from the current phase relation in the similar device is around 0.9 with the junction length of 40 nm (Our JJ2 length is 100 nm which is longer than this. Then $\tau$ in our device can be smaller than 0.9. This is justified in the green curve of Fig. S6(a) where the oscillation reflecting the current phase relation of JJ1 is skewed but not so significant. When we fit this with the current phase relation equation, we can get $\tau \sim 0.7$). With the critical current of JJ2 is 200 nA, $\frac{dI_{\text{JJ2}}}{d\phi}(\phi_{20} = \pi) = -0.416$ μA is obtained with $\tau = 0.9$.

Then, we need to evaluate the inductance of our SC arms. The kinetic inductance can be estimated from the arm geometry and resistivity with

$$L_K = \frac{l}{w}\frac{h}{2\pi^2}\frac{R_n}{\Delta} \qquad \text{Eq. 3}$$

, where $l, w, R_n,$ and $\Delta$ are the length, width, resistivity in the normal state, and SC gap energy, respectively [3]. For the resistivity, we use 6.4 Ω, as has been used for the same quantum well structure described in the literature [3]. Then, $L_{\text{arm1}} = 51.8$ pH is determined as the kinetic inductance of the central arm. Using this result, the phase shift per 1 μA is estimated as $2\pi L_{\text{arm1}}/\Phi_0 = 0.05\pi/\mu A$. In addition, $2\pi L_{\text{arm2}}/\Phi_0 = 0.15\pi/\mu A$ is also calculated.

Therefore, using the estimated $\frac{dI_{\text{JJ2}}}{d\phi}$ and the inductances,

$$0 < \frac{L_{\text{arm1}} I_1}{\frac{\Phi_0}{2\pi} + (L_{\text{arm1}} + L_{\text{arm2}})\frac{dI_{\text{JJ2}}}{d\phi}(\phi_{20}=0)} \leq \Delta\phi_2 \leq \frac{L_{\text{arm1}} I_1}{\frac{\Phi_0}{2\pi} + (L_{\text{arm1}} + L_{\text{arm2}})\frac{dI_{\text{JJ2}}}{d\phi}(\phi_{20}=\pi)} = 0.068\pi I_1 * 10^6 \qquad \text{Eq. 4}$$

is obtained in the case of $I_1 > 0$ because of $\frac{dI_{\text{JJ2}}}{d\phi}(\phi_{20} = \pi) \leq \frac{dI_{\text{JJ2}}}{d\phi}(\phi_{20}) \leq \frac{dI_{\text{JJ2}}}{d\phi}(\phi_{20} = 0)$ and $\frac{2\pi}{\Phi_0}(L_{\text{arm1}} + L_{\text{arm2}})\frac{dI_{\text{JJ2}}}{d\phi}(\phi_{20} = \pi) > -1$. In the case of $I_1 < 0$, all the inequality signs in Eq. 4 are reversed.

We consider converting $\Delta\phi_2$ induced by $I_1$ to the change of $B$. If there is no phase shift due to $I_1$, the results can be written as $I_{\text{sw+}}(B_i)$ with $B_i = B_0(\phi_{20}/2\pi - (L_{\text{arm1}} - L_{\text{arm2}})I_{\text{cw}}/\Phi_0)$. Here $B_0$ gives the external magnetic field producing the flux quanta in the SC loop. When the inductance effect due to $I_1$ is included, the obtained switching current curve is written as $I_{\text{sw+}}(B_f)$ with $B_f = B_0((\phi_{20} + \Delta\phi_2)/2\pi - (L_{\text{arm1}} - L_{\text{arm2}})I_{\text{cw}}/\Phi_0) = B_i + B_0\Delta\phi_2/2\pi$. This indicates that the inductance effect induced by $I_1$ shifts the $I_{\text{sw+}}$ curve to the negative direction by $B_0\Delta\phi_2/2\pi$ along the $B$ axis, resulting in generating the SDE.

Herein, we remove the inductance effect from $I_1$ by shifting the observed $I_{sw+}(B)$ to the positive direction. From Eq. 4, $0 < B_0 \Delta\phi_2/2\pi \leq 0.034 B_0 I_{sw+} * 10^6$ is obtained in the case of $I_{sw+} > 0$. On the other hand, the observed $I_{sw-}(B)$ need to be shifted to the negative direction in the case of $I_{sw-} > 0$ with $|B_0 \Delta\phi_2|/2\pi$ and $0.034 B_0 I_{sw-} * 10^6 < B_0 \Delta\phi_2/2\pi \leq 0$. $B_0$ can be evaluated from the single period of the switching current oscillation as a function of $B$ in Fig. 2(c). Then we remove the inductance effect from Fig. 2(c) by shifting the curves with $0.034 B_0 I_{sw+} * 10^6$, resulting in Fig. S6(c). The SDE regions still exist in Fig. S6(c). Herein we can conclude that the SDE induced by the inductance is negligible.

We note that the obtained SDE region around the local minimum of the switching current oscillation almost disappears after the correction with $L_{arm1} = 113$ pH and $L_{arm2} \simeq 3 L_{arm1}$ as shown in Fig. S6(f). However, in the case, the SDE does not completely disappear from the switching current oscillation. Figure S6(g) indicates that the SDE regions appear around the local maximum of the switching current oscillation. Therefore, it is impossible to kill the SDE from the oscillation only with the correction of Eq. 4.

To check the evaluated inductance is valid, we utilize the switching current measurement of the SQUID consisting of JJ1 and LJJ1 with JJ2 pinched off and LJJ2 on with the bias current $I_1$. The results are shown in Fig. S6(a) as the green curve. The oscillation reflects the current phase relation of JJ1. Then we apply additional current $I_2$.

From the discussion about the setup of Fig. S5(a) and Eq.2, $\Delta\phi_2 \simeq 2\pi L_{arm1} I_1/\Phi_0$ is obtained when $\frac{dI_{JJ2}}{d\phi}(\phi_{20}) \simeq 0$. This means that when the phase difference of JJ2 ($\phi_{20}$) gives the critical current of JJ2, $\Delta I_{cw}$ in Fig. S5(a) equals to zero.

This conclusion can be applied in the setup of Fig. S5(c). With no $I_2$, the bias current of $I_1$ is used to detect the switching current of the JJ1 and LJJ1. When the additional current of $I_2$ is applied, $I_2$ distributes to the SC arm 3 and the other SC arm including LJJ1 and JJ1. Around the magnetic fields which produce the maximum of the switching current oscillation as the green curve in Fig. S6(a), we can assume that all $I_2$ flows in the SC arm 3 because the maximum point of the switching current oscillation of the SQUID gives the phase difference in JJ1 generating the JJ1 critical current. In this case, the oscillation pattern shifts due to $2\pi L_{arm3} I_2/\Phi_0$.

The obtained interference patterns with $I_2 = \pm 0.75$ and $\pm 1$ µA are indicated in Fig. 5(a). As expected, the opposite current produces an opposite shift along the $B$ axis.

We then evaluate the shifts of the two interference patterns ($\delta B$) obtained with the same $|I_2|$ and plotted them as a function of $\delta I_2 = 2|I_2|$ in Fig. S6(b). We measure the shift for $I_2 = \pm 0.625, \pm 0.75, \pm 0.875,$ and $\pm 1$ µA. Then, $\delta B$ can be converted to the phase shift from the inductance effect divided by the interference period. Consequently, the phase shift induced by the inductance effect per 1 µA of $I_2$ can be evaluated as $2\pi L_{arm3}/\Phi_0 = 0.0805\pi/µA$ from the tilt of the line obtained using proportional fitting in Fig. S6(b). The geometry of the SC arm 3 is rectangular with a 0.5 µm width and 4 μm length. In the SDE measurement, the inductance effect is induced by the central SC arm 1 (0.35 μm width, 3 μm length). In our case, the dominant inductance contribution is derived from the kinetic inductance following Eq. 3. Therefore, the central SC arm 1 produces $2\pi L_{\text{arm1}}/\Phi_0 = 0.0863\pi/\mu A$. This is approximately consistent with $2\pi L_{\text{arm1}}/\Phi_0 = 0.05\pi/\mu A$ calculated from Eq. 3.

We also measure an asymmetric SQUID device of JJ2 and LJJ2 in the same structure as that of the device we use in the main manuscript. The results of the positive (red) and negative (blue) switching currents are shown in Fig. S6(h). We enlarge the results around the maximum and minimum currents with the proper offsets to compare the magnetic fields producing the maximum of the positive current oscillation and minimum of the negative current oscillation in Fig. S6(i). As arrowed in the figure, the maximum and minimum points appear at the same $B$ with the resolution of 0.006 mT. This means that the inductance of the SC loop is at least lower than 64 pF. This result also supports that our inductance evaluation is valid.

**Supplementary Note 5: Measurement of JJ1 and JJ2 with three-terminal configuration**

It is important to confirm that LJJ2 sufficiently shunts JJ2 in the asymmetric SQUID and to study the effects caused by tuning the asymmetry of LJJ2 and JJ2. For this purpose, three-terminal measurements are performed. Three-terminal measurements have been discussed with respect to the multi-terminal Josephson effect [4–6]. In this configuration, the bias currents $I_1$ and $I_2$ are applied to the device, and $V_1$ and $V_2$ are measured. We then explore the three-terminal measurements on our device with JJ1, JJ2, and LJJ2 on. In this case, $V_2$ should be zero if LJJ2 sufficiently shunts the two SC electrodes associated with JJ2. The obtained results for $V_1$ and $V_2$ at $B = 0.012$ mT are shown in Figs. S7(a) and (b), respectively, and as expected, indicate $V_2 \approx 0$ mV

When the gate voltage on LJJ2 ($V_{\text{gL2}}$) becomes negative, the shunt between the two SC electrodes of JJ2 becomes weak, and $I_2$ can also induce dissipation in JJ1, as shown in Figs. S8(a)–(f). At $V_{\text{gL2}} \approx -1.6$ V, $V_2$ is still zero in the measurement range, whereas $V_2$ becomes finite at $V_{\text{gL2}} \approx -1.8$ V. This result implies that the shunt induced by LJJ2 is broken when $V_{\text{gL2}} \lesssim -1.8$ V. The switching current of LJJ2 becomes equivalent to that of JJ2 at $V_{\text{gL2}} \approx -1.7$ V. In this situation, a change in $I_1$ also changes $\phi_2$ because of the less shunt on LJJ2. This can be found as the result that the switching current of the SQUID involving JJ2 and LJJ2 is modulated by $I_1$ as shown in Figs. S8(d) and (f). At $V_{\text{gL2}} = -2.0$ V, the circular critical current contour[5] of the hybridized JJ1 and JJ2 can be obtained, as shown in Figs. S8(e) and (f). Fig. S7 does not indicate such a circular critical current contour, indicating that LJJ2 strongly shunts the two SC electrodes of JJ2 and that the phase induced by the magnetic flux in the loop shifts the phase difference of JJ2.

Figures S9(a)–(f) show $V_1$ and $V_2$ measured when $V_{\text{gL2}} = -1.6$ V, $-1.8$ V, and $-2.0$ V. At $V_{\text{gL2}} \lesssim -1.8$ V, the switching current oscillation amplitude decreases because the circular supercurrent in the loop is restricted by the LJJ2 supercurrent. In addition, the SDE sign at $V_{\text{gL2}} \lesssim -1.8$ V is reversed compared to that at $V_{\text{gL2}} = -1.6$ V. This occurs because $I_1$ also shifts $\phi_2$ as discussed in the previous paragraph. Therefore, $V_2$ becomes finite at $V_{\text{gL2}} \lesssim -1.8$ V while $V_2$ is almost zero at $V_{gL2} \simeq 0$ mV. In the regime of $V_{gL2} \lesssim -1.8$ V, it is difficult to analyze the results quantitatively because we cannot estimate $\phi_2$, unlike the $V_{gL2} \gtrsim -1.6$ V case where $\phi_2 \simeq 2\pi\Phi/\Phi_0$ is utilized. $\Phi$ and $\Phi_0$ represent the magnetic flux in the loop and the magnetic flux quantum, respectively.

**Supplementary Note 6: Numerical model and calculation**

The numerical calculation conducted in this study is described in this section. We use the tight-binding model that discretizes the real space on a square lattice. The three-terminal structure describing the two coupled JJs is shown in Fig. S10(a). The three yellow-shadowed regions are the SC regions, where the SC pair potentials $\Delta_\alpha = \Delta_0 e^{i\phi_\alpha}$ ($\alpha$ = 1, C, 2) represent the SC proximity effect in the InAs quantum well from the Al electrodes. The two normal regions are displayed in white. Our model is the SNSNS junction which corresponds to JJ1 and JJ2 in our device.

First, we explain briefly how this model and our calculation correspond to our experimental results. In our measurement, we detect the switching current of JJ1 with JJ2 in the lower SC loop with LJJ1 pinched off and LJJ2 on. We always pinch off the LJJ1, meaning that we can ignore the interference from the upper SC loop. We note that the phase difference of LJJ2 included in the lower loop is ignorable because the switching current of LJJ2 is much larger than that of JJ2. Then the $B$ dependence of $I_{sw1}$ can be written as $I_{sw1}(\phi_2)$ because $B$ changes the magnetic flux in the lower loop, changing $\phi_2$. To obtain the switching current of JJ1 as a function of $\phi_2$, in our below calculation using the model, we calculate the current phase relation of JJ1 as a function of $\phi_1$ at the fixed $\phi_2$ and then we obtain the positive and negative JJ1 critical currents at the fixed $\phi_2$ from the current phase relation as the maximum and minimum supercurrents, respectively. Therefore, the calculation based on the SNSNS model can be applied to reproduce our experimental results.

In this model, we evaluate the Josephson current using the phase differential of the free energy $E_J$ as

$$I_\alpha(\phi_1, \phi_2) = -\frac{e}{\hbar}\frac{\partial E_J(\phi_1, \phi_2)}{\partial \phi_\alpha}$$

with ($\alpha$ = 1, 2). Here, we set $\phi_c = 0$. The positive and negative critical currents are the maximum and minimum values of the supercurrent in JJ1 ($I_{s1}(\phi_1, \phi_2)$) when $\phi_1$ is swept. The parameters for the simulation are as follows. The various lengths denoted in Fig. S10(a) are $L_w = 21a$, $L_c = 13a$, $L_N = 6a$, and $L_s = 41a$ with lattice constant $a = 20$ nm for the tight-binding model. The Fermi energy is set as $E_F = 0.9t$ and the pair potential is $\Delta_0 = 0.03t$ with the energy scale $t = \hbar^2/(2m^*a^2) \simeq 4.141$ meV. The coherent length is evaluated as $\xi \simeq 240$ nm if impurities are absent. For InAs, the effective mass is $m^* = 0.023 m_e$. At the sites on the boundaries of the SC regions, the strength of the pair potential is divided by two. We provide on-site random potentials $U(i,j)$ to describe the effects of impurities. The range of randomness is $W_0/2 \geq U(i,j) \geq -W_0/2$ with $W_0 = 2t$, which corresponds to a mean free path of $l \simeq 217$ nm. The junction energy is evaluated by the summation of the Andreev levels, $E_J(\phi_1, \phi_2) = \sum_n E_n(\phi_1, \phi_2)$. For the short junction case, we consider levels up to $n = 12$ for positive energy, including the spin degrees of freedom in the calculations.

We perform a numerical simulation using the tight-binding model for the critical current, which is the absolute value of either the maximum or minimum of the current phase relation when JJ1 shares an SC electrode with JJ2. Figure S10(b) shows the numerical results of the positive and negative critical currents in JJ1 ($I_{c+}$ and $I_{c-}$) as a function of $\phi_2$. The critical current oscillates periodically as a function of $\phi_2$, reflecting the nonlocal Josephson Effect originating from the coherent coupling between the two JJs. The maximum and minimum points in the oscillation are slightly shifted from $\phi_2 = 0, \pi$. However, the difference is more notable around $\phi_2 = \pm\pi$ due to a sharp change in $I_{c+}$ and $I_{c-}$, which results in a large SDE, as shown in Fig. S10(c). The SDE sign is reversed when $\phi_2$ passes $\pm\pi$. This result is in good agreement with the experimentally measured features shown in Fig. 2(c) and further supports the conclusion that the observed SDE is derived from the coherent coupling between JJ1 and JJ2. The SDE also appears in the vicinity of $\phi_2 \sim 0$ in the numerical results. In our experiments, these SDE regions are not observed systematically. Figure 2(c) indicates that such small SDE regions exist but cannot be observed clearly using the device discussed in Supplementary Note 7 or the mirrored structure discussed in Supplementary Note 3. A possible reason for this is the inductance effect, which causes a shift along the *B* axis dependent on the current direction that could easily erase the small SDE, or the resolution of our measurement setup may be insufficient.

**Supplementary Note 7: SDE reproducibility using a different device**

We check the reproducibility of the SDE using a different device we refer to as #B. Figure S11(a) shows the scanning electron microscopic image and schematic image of device #B. The loop area is reduced to 7.8 μm² compared to the 11.7 μm² in the device used for the main manuscript. Figure S11(b) shows the measured V-I of JJU with different sweep directions of bias current *I*. There is no clear hysteresis, indicating that JJU is overdamped. Figures S11(c), (d), and (e) show *V* as a function of *I* and the gate voltages for the single JJU, JJL, and LJJ1 of the larger JJ in the loop, respectively.

Then we study the SDE in this device. Figure S12(a) presents *V* as a function of *I* and *B*. When we choose the line profiles at *B* = 0.100 mT and 0.085 mT, as shown in Fig. S12(b), the SDE is clearly observed. Then, $|I_{sw+}|$ and $|I_{sw-}|$ as a function of *B* are shown in Fig. S12(c) using red and blue circles, respectively. The results clearly reproduce the SDE in the vicinity of the phase difference on JJL $\phi(B) = \pi$, as obtained in Fig. 2(c).

**Supplementary Note 8: No SDE in a device with a wider separation between JJ1 and JJ2**

We observe the switching current oscillation and SDE in the coupled JJ device of JJ1 and JJ2, and conclude that the coherence of the coupling is essential for this observation. This is because the oscillation of the JJ1 switching current is only obtained when the separation of two JJs is shorter than the coherent length of the SC electrode as already discussed in the supplementary information of our previous study [7]. To check if the SDE vanishes in the device where the separation of two JJs is comparable to or longer than the coherence length of Aluminium (1 μm), we fabricated the device whose optical microscope image is shown in Fig. S13(a). The JJ1 is out of the SC loop while the JJ2 is in the loop. There is no larger JJ in the SC loop. The designed separation of JJ1 and JJ2 is 1 um which is longer than that of the device used in the main manuscript. The measured voltage $V$ of JJ1 as a function of $I$ and $B$ is shown in Fig. S13(b). The white region corresponds to the supercurrent region. No oscillation can be seen and therefore, no SDE appears. This additional experiment can further prove the necessity of the coherence of the coupling.

**Supplementary Note 9: Comparing to the multiterminal JJ results**

The multiterminal Josephson junctions are defined as systems having a single normal metal connected to multiple SC electrodes. In such cases, the Andreev bound states in the normal metal depend on more than 3 SC phases. Using the multiterminal JJ structures, some groups claim observation of the SDE and nonreciprocal supercurrent related to the symmetry breaking of the multiterminal JJs. In the experimental reports, the multiterminal JJs are biased with more than two current sources ($I_1$ and $I_2$, for example) and the SDE is obtained as the different absolute values of the positive and negative switching currents about $I_1$ with the fixed $I_2$. However, when more than two current sources are allowed to be used, such results can be produced even in the measurement of the single two-terminal JJs. As an example, we measure the two-terminal JJ fabricated from the same wafer we used in the experiments on the main manuscript with the setup depicted in Fig. S14(a). The measured result of the voltage difference, $V$ as a function of $I_1$ and $I_2$ at no magnetic field is shown in Fig. S14(b). The line profiles at $I_2 = -0.5, 0, 0.5$ μA are exhibited as the red, green, and blue, respectively. The finite $I_2$ can shift the $V$ curve as a function of $I_1$, resulting in the nonreciprocal supercurrent as a function of $I_1$. However, this does not mean that the time-reversal and spatial-inversion symmetries in the single two-terminal JJ are broken. As represented by this example, it is difficult to discuss and reveal the SDE origin and the symmetry breaking in the JJs with more than two current sources.

**References**

[1] M. Kjaergaard, H. J. Suominen, M. P. Nowak, A. R. Akhmerov, J. Shabani, C. J. Palmstrøm, F. Nichele, and C. M. Marcus, *Transparent Semiconductor-Superconductor Interface and Induced Gap in an Epitaxial Heterostructure Josephson Junction*, Phys. Rev. Applied **7**, 034029 (2017).

[2] G. Nanda, J. L. Aguilera-Servin, P. Rakyta, A. Kormányos, R. Kleiner, D. Koelle, K. Watanabe, T. Taniguchi, L. M. K. Vandersypen, and S. Goswami, *Current-Phase Relation of Ballistic Graphene Josephson Junctions*, Nano Lett. **17**, 3396 (2017).

[3] F. Nichele et al., *Relating Andreev Bound States and Supercurrents in Hybrid Josephson Junctions*, Physical Review Letters **124**, 226801 (2020).

[4] A. W. Draelos, M.-T. Wei, A. Seredinski, H. Li, Y. Mehta, K. Watanabe, T. Taniguchi, I. V.

Borzenets, F. Amet, and G. Finkelstein, *Supercurrent Flow in Multiterminal Graphene Josephson Junctions*, Nano Letters **19**, 1039 (2019).

[5] N. Pankratova, H. Lee, R. Kuzmin, K. Wickramasinghe, W. Mayer, J. Yuan, M. G. Vavilov, J. Shabani, and V. E. Manucharyan, *Multiterminal Josephson Effect*, Physical Review X **10**, (2020).

[6] G. V. Graziano, J. S. Lee, M. Pendharkar, C. J. Palmstrøm, and V. S. Pribiag, *Transport Studies in a Gate-Tunable Three-Terminal Josephson Junction*, Physical Review B **101**, (2020).

[7] S. Matsuo, J. S. Lee, C.-Y. Chang, Y. Sato, K. Ueda, C. J. Palmstrøm, and S. Tarucha, *Observation of Nonlocal Josephson Effect on Double InAs Nanowires*, Commun Phys **5**, 221 (2022).

**Supplementary Figures**

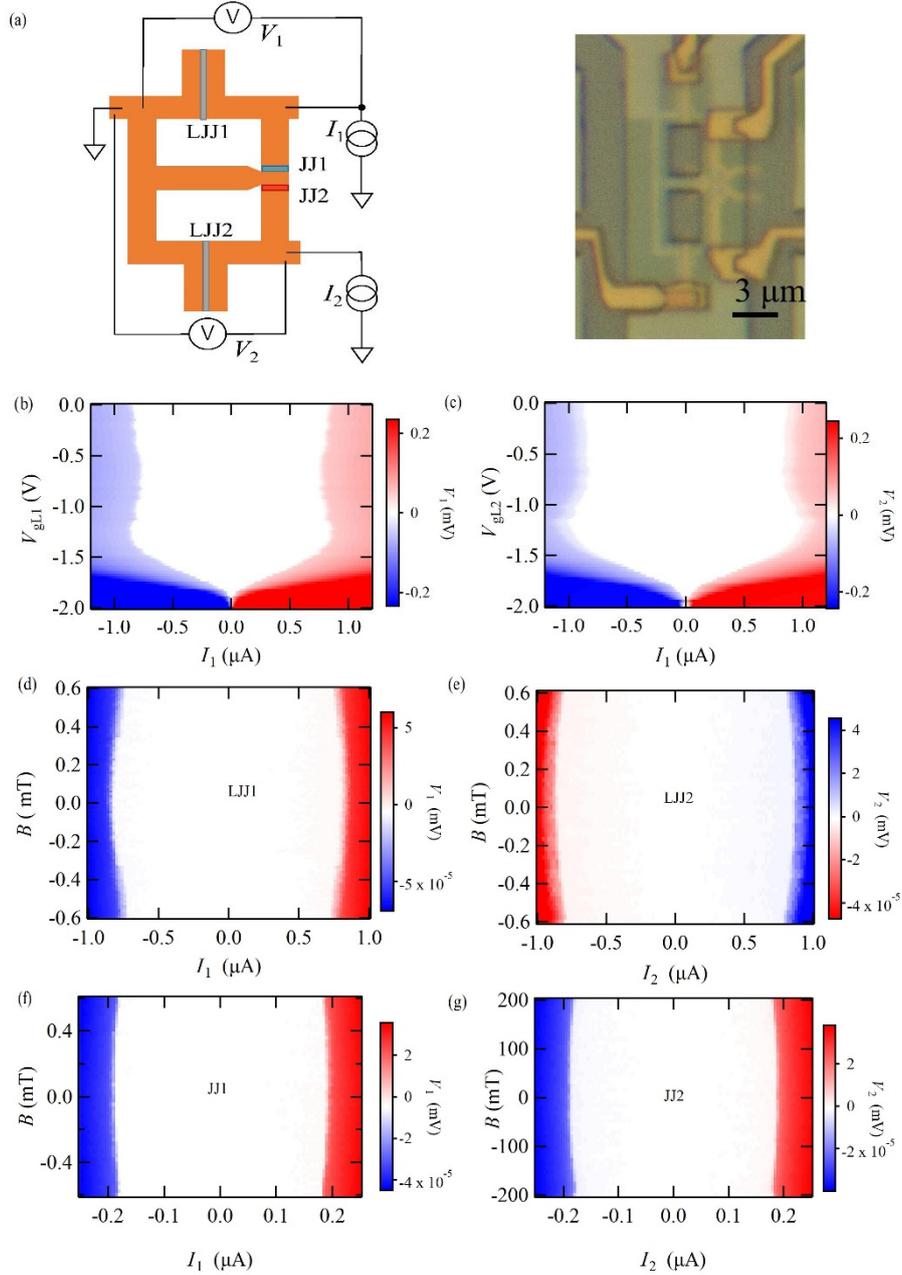

**Fig. S1**

(a) Left: Entire device schematic and measurement setup. Right: Optical image of our dummy device. (b) $V_1$ as a function of $I_1$ and $V_{gL1}$ in the single LJJ1 case with the other JJs pinched off. (c) $V_2$ as a function of $I_2$ and $V_{gL2}$ in the single LJJ2 case with the other JJs pinched off. (d), (e), (f), and (g) indicate the magnetic field dependence of the single LJJ1, LJJ2, JJ1, and JJ2, respectively. Our measurement range is much narrower than the period of the Fraunhofer pattern, resulting in less dependence on $B$ by the switching current in the single JJs.

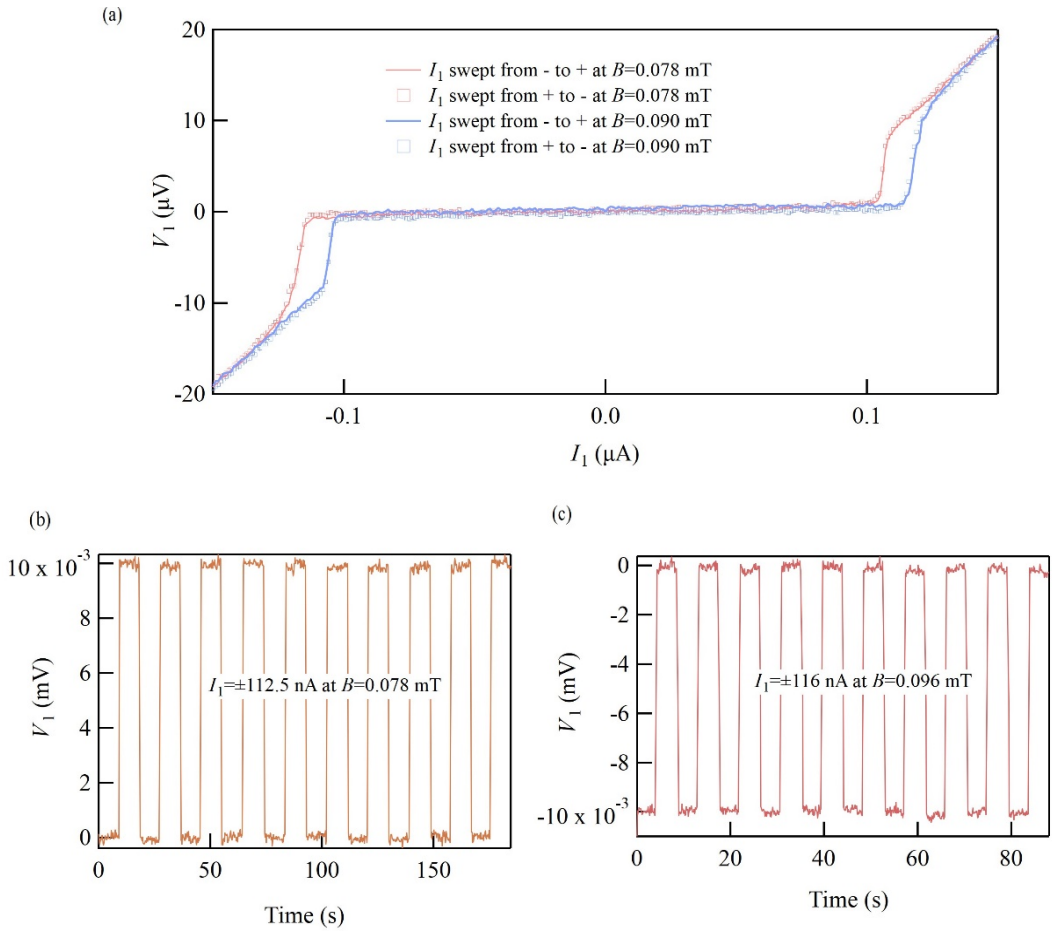

**Fig. S2**

(a) $V_1$ as a function of $I_1$ at $B = 0.078$ and $0.090$ mT shown by the red and blue lines and squares, respectively. The lines and squares indicate positive and negative current sweep directions, respectively. (b) $V_1$ obtained for $B = 0.078$ mT when applying $I_1 = -112.5$ nA for 9.1 s and then $I_1 = 112.5$ nA for 9.1 s ten times. (c) $V_1$ at $B = 0.096$ mT when applying $I_1 = -116$ nA for 4.4 s and then $I_1 = 116$ nA for 4.4 s ten times.

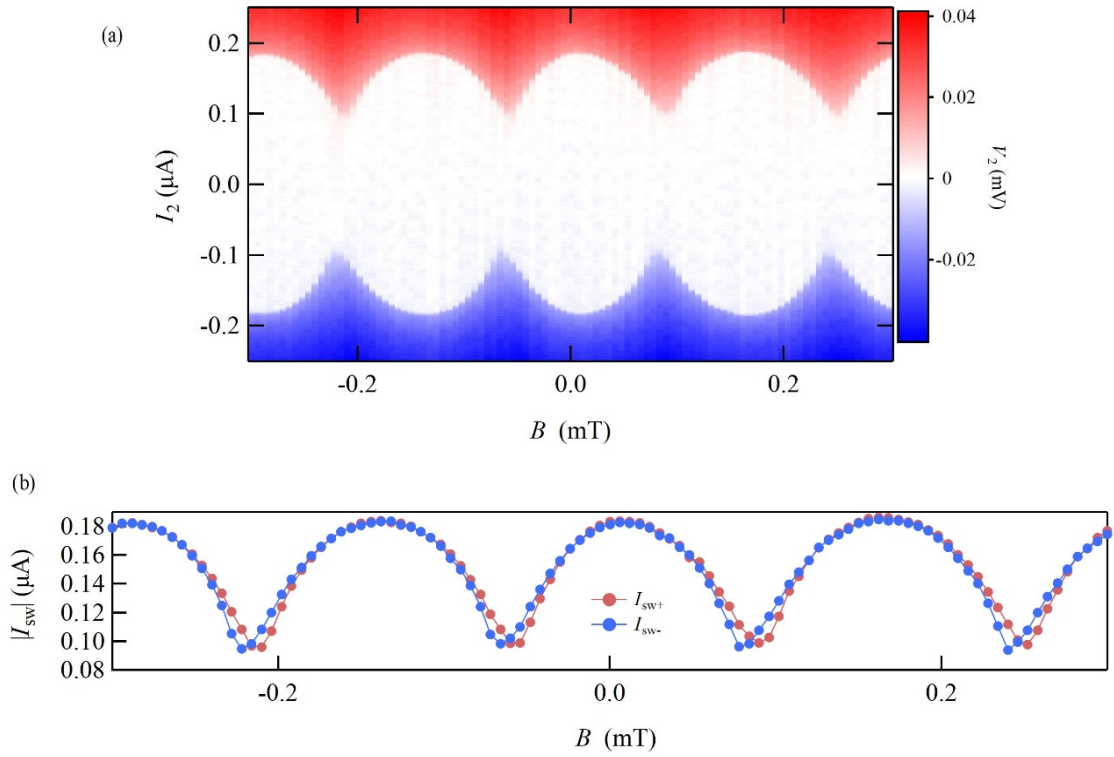

**Fig. S3**
(a) $V_2$ as a function of $I_2$ and $B$ with LJJ2 pinched off. The results are for the mirrored structure.
(b) $I_{sw+}$ and $I_{sw-}$ represented as the red and blue circles, respectively. The SDE is observed in the vicinity of the minimum $|I_{sw}|$.

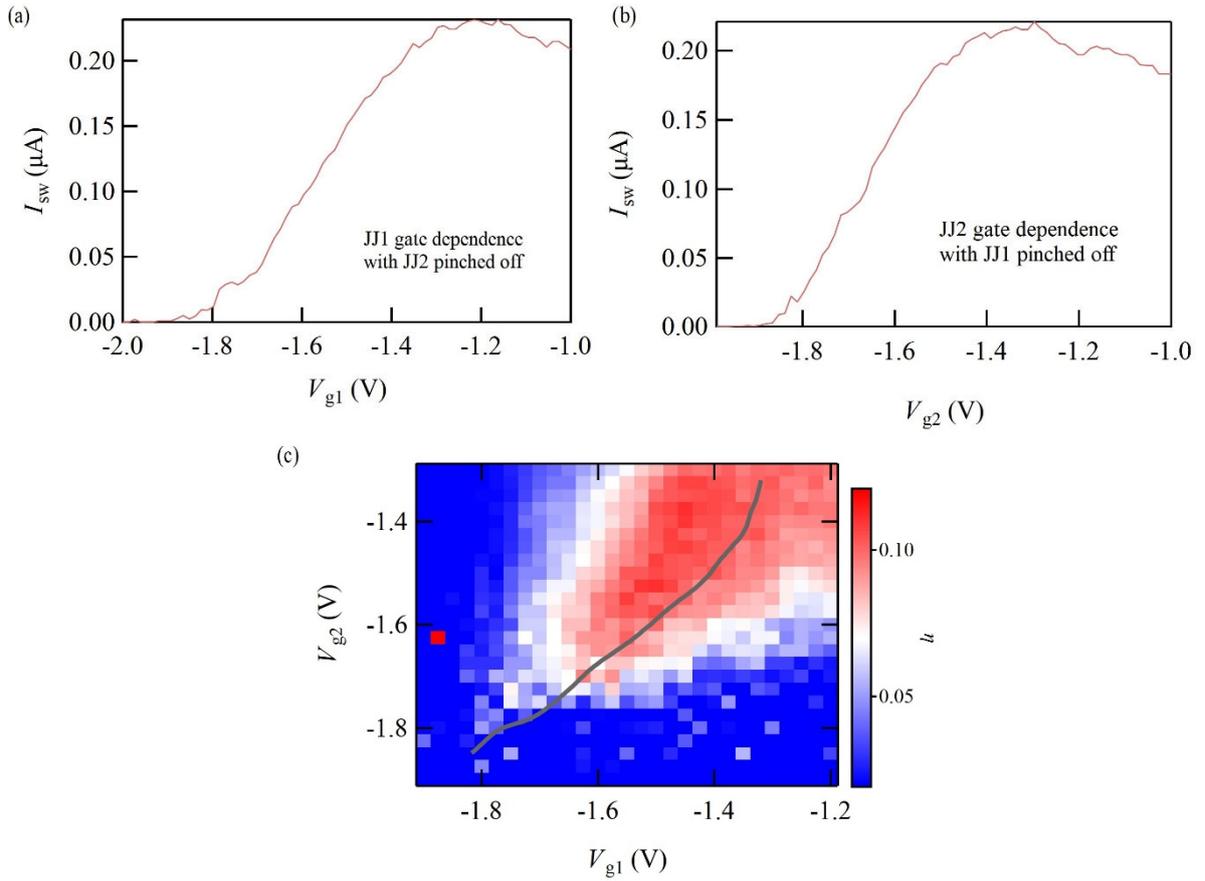

**Fig. S4**

(a) and (b) The switching current dependence on $V_{g1}$ and $V_{g2}$ in the single JJ1 and JJ2 cases, respectively, after we changed the cables to use the mirrored structure configuration. (c) The SDE ratio η as a function of $V_{g1}$ and $V_{g2}$ at $B = 0.0768$ mT. The grey curve exhibits ($V_{g1}, V_{g2}$) producing the same switching currents in JJ1 and JJ2 evaluated from (a) and (b). The diagonal red SDE region follows the grey curve, as is also observed in Fig. 3(d).

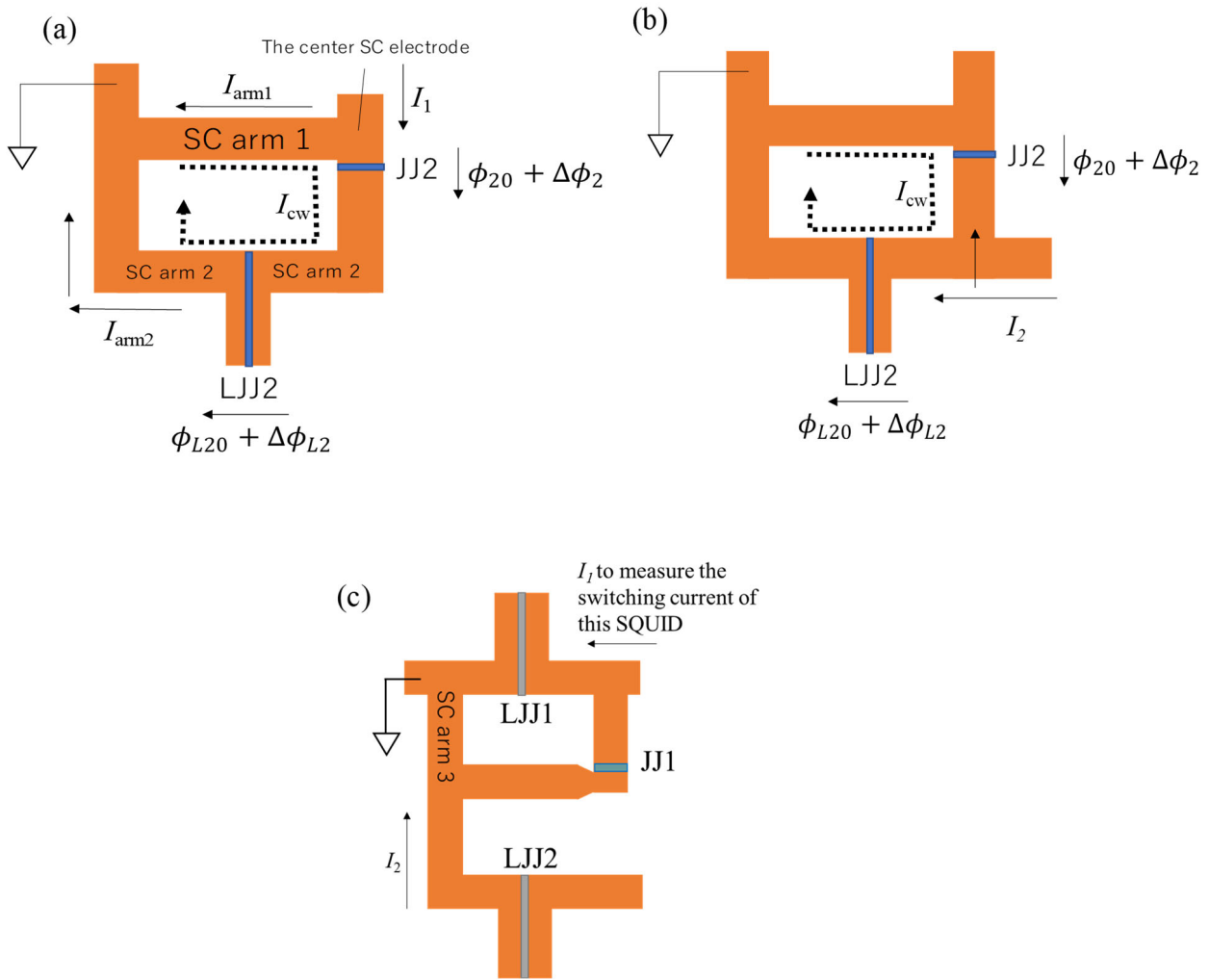

**Fig. S5**
(a) A schematic image of the device below the center SC arm that JJ1 and JJ2 share. When measuring JJ1, the bias current $I_1$ flows in either the SC arm 1 or SC arm 2 as labeled in this figure to the GND.
(b) A schematic image when the switching current of the SQUID of JJ2 and LJJ2 is measured. The bias current of $I_2$ flows through either JJ2 or LJJ2 to the GND.
(c) A schematic image when we roughly estimate the inductance of the SC arm 3 with JJ2 pinched off. To measure the switching current of the SQUID, we apply $I_1$ and measure the voltage difference. Then we apply the additional $I_2$ to induce the phase shift through the inductance of the SC arm 3.

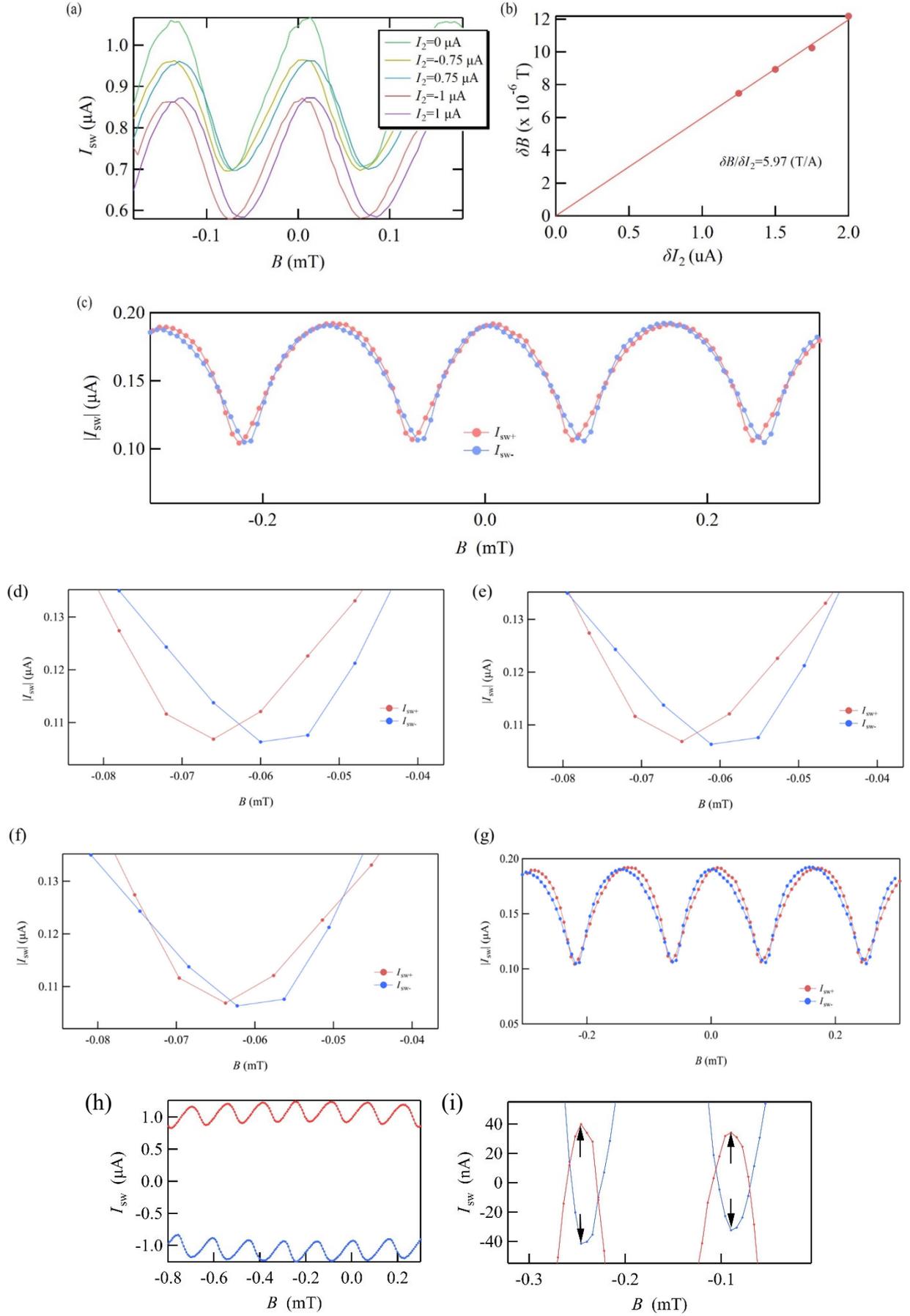

**Fig. S6**

(a) Interference pattern obtained during measurement of the SQUID formed by JJ1 and LJJ1 with $I_2 = 0, \pm 0.75$ µA and $\pm 1$ µA when JJ2 is pinched off. The additional current flowing in the left arm of the SQUID induces the shift along the $B$ axis. (b) The obtained shift $\delta B$ along the $B$ axis as a function of $\delta I_2$ is shown by the red circles. The solid line indicates the fitted result using proportionality. (c) $|I_{sw}|$ as a function of $B$ corrected by the inductance effect evaluated from (a) and (b). The inductance causes a negligible shift of the oscillation along the $B$ axis. (d) The enlarged SDE region in Fig. 2(c). (e) The positive and negative switching current curves in the absolute value including the correction with $L_{arm1} = 80$ pH, $L_{arm2} \simeq 3L_{arm1}$. (f) and (g) represents the positive and negative switching current curves in the absolute value including the correction with $L_{arm1} = 113$ pH, $L_{arm2} \simeq 3L_{arm1}$.

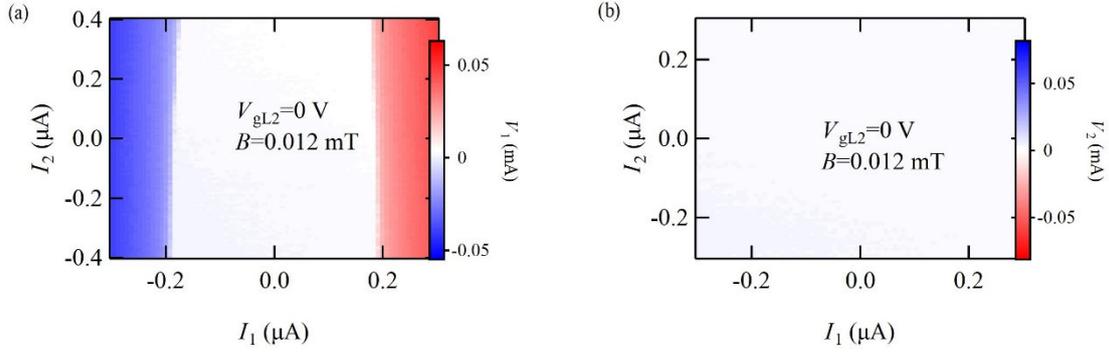

**Fig. S7**

(a) and (b) show $V_1$ and $V_2$ as a function of $I_1$ and $I_2$ at $B = 0.012$ mT with LJJ1 pinched off, respectively.

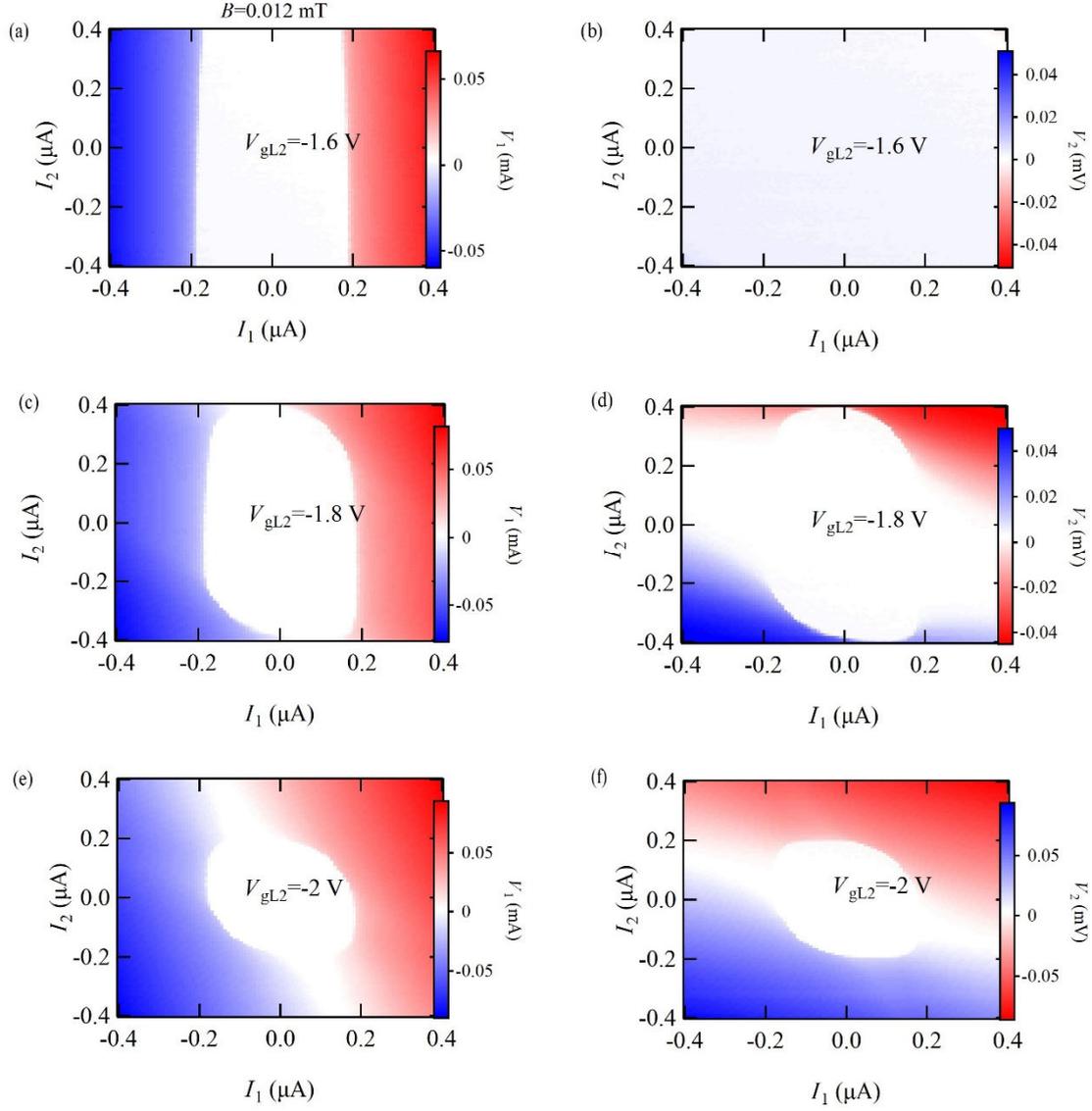

**Fig. S8**

(a) and (b) show $V_1$ and $V_2$ as a function of $I_1$ and $I_2$ at $B = 0.012$ mT with LJJ1 pinched off at $V_{gL2} = -1.6$ V, respectively. (c) and (d) show $V_1$ and $V_2$ as a function of $I_1$ and $I_2$ at $B = 0.012$ mT with LJJ1 pinched off at $V_{gL2} = -1.8$ V, respectively. (e) and (f) show $V_1$ and $V_2$ as a function of $I_1$ and $I_2$ at $B = 0.012$ mT with LJJ1 pinched off at $V_{gL2} = -2.0$ V, respectively.

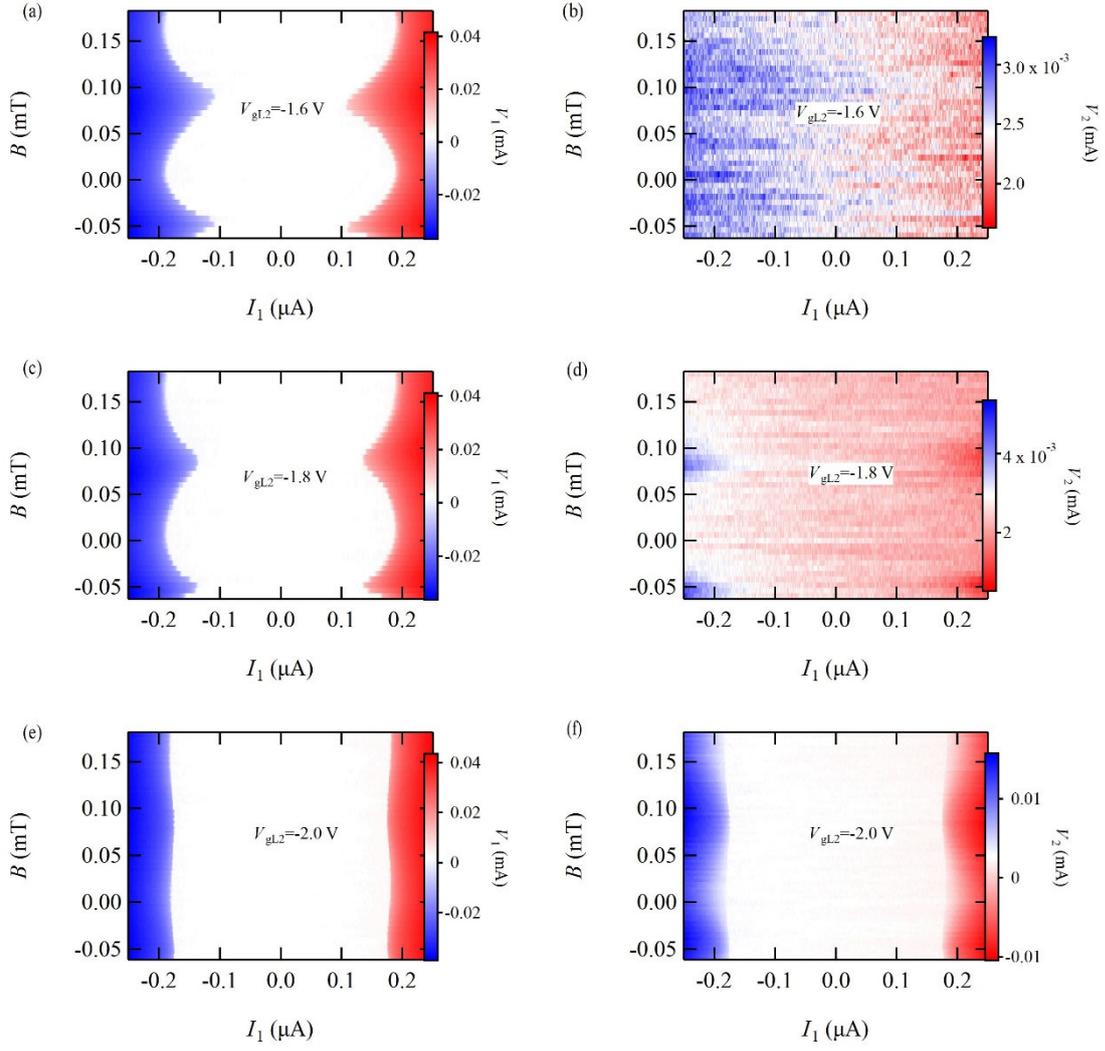

**Fig. S9**

(a) and (b) show $V_1$ and $V_2$ as a function of $I_1$ and $B$ with LJJ1 pinched off at $V_{gL2} = -1.6$ V, respectively. (c) and (d) show $V_1$ and $V_2$ as a function of $I_1$ and $B$ with LJJ1 pinched off at $V_{gL2} = -1.8$ V, respectively. (e) and (f) show $V_1$ and $V_2$ as a function of $I_1$ and $B$ with LJJ1 pinched off at $V_{gL2} = -2.0$ V, respectively.

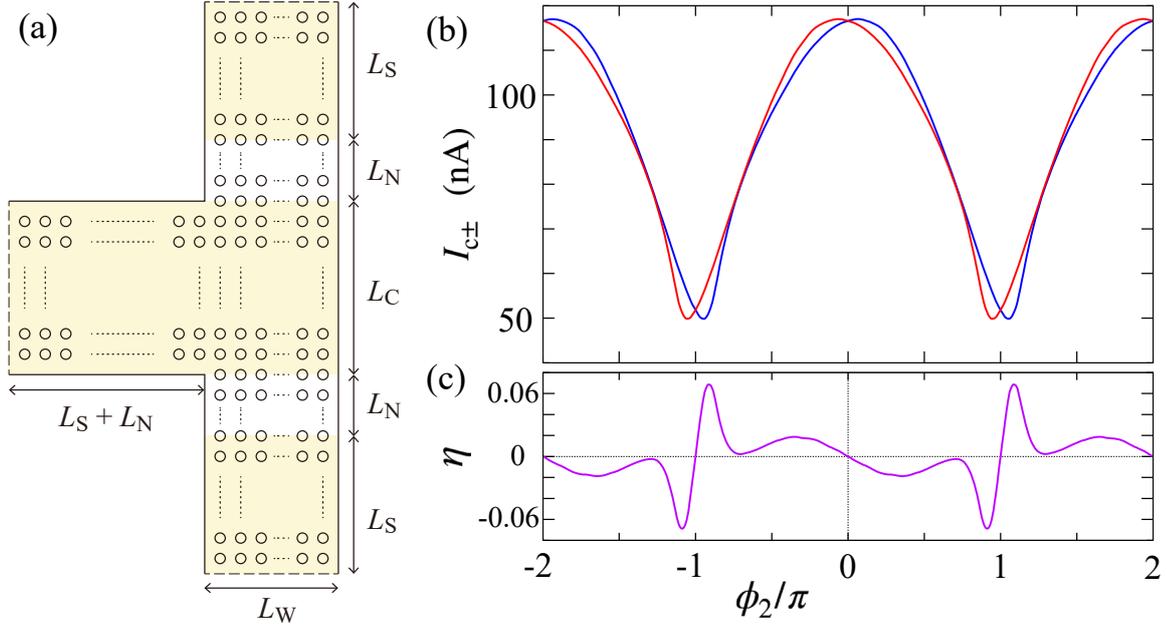

**Fig. S10**
(a) Tight-binding model for coupled JJs described in two-dimensional space, where $L_W, L_C, L_N$ and $L_S$ are the widths of the upper and lower SC terminals, the width of the center SC terminal, the length of normal regions, and the length of upper and lower SC regions, respectively, with lattice constant $a$. (b) Absolute value of critical current in positive (red, $I_{c+}$) and negative (blue line, $I_{c-}$) directions when the nonlocal phase $\phi_2$ is tuned. (c) Coefficient for the SDE, $\eta = (I_{c+} - I_{c-})/(I_{c+} + I_{c-})$.

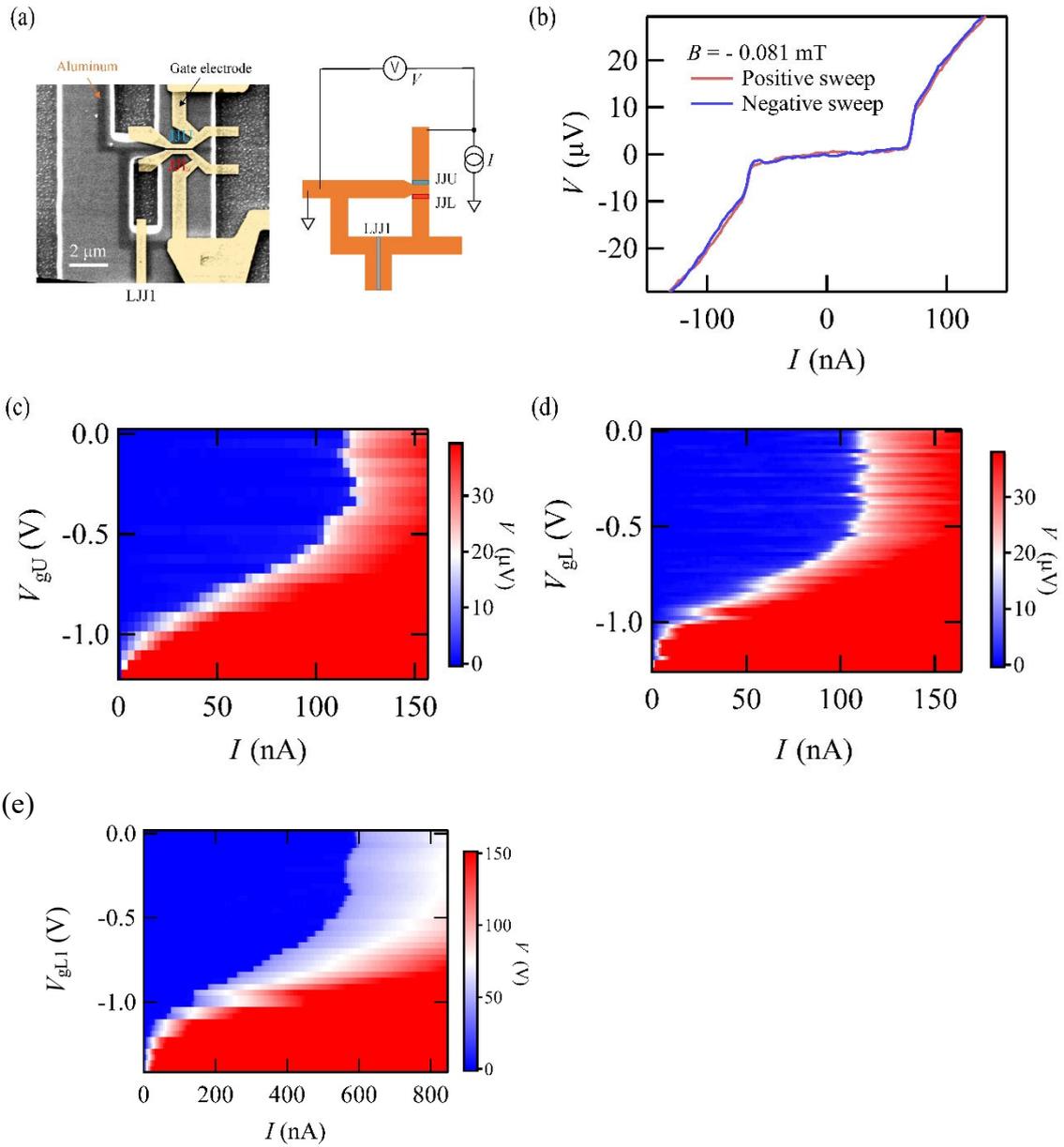

**Fig. S11**

(a) Scanning electron microscopic image and schematic of device #B. Two coupled JJs, denoted by JJU and JJL, are prepared and JJL is embedded in the SC loop containing a larger JJ denoted by LJJ1. The phase difference in JJL is controlled by the out-of-plane magnetic field. (b) Measured voltage $V$ as a function of $I$ in the JJU measurement with an opposite sweep direction at $B = -0.081$ mT. No hysteresis is observed, indicating that JJU is overdamped. (c), (d), and (e) show $V$ as a function of $I$ and gate voltages measured for the single JJU, JJL, and LJJ1 cases with the other JJs pinched off, respectively.

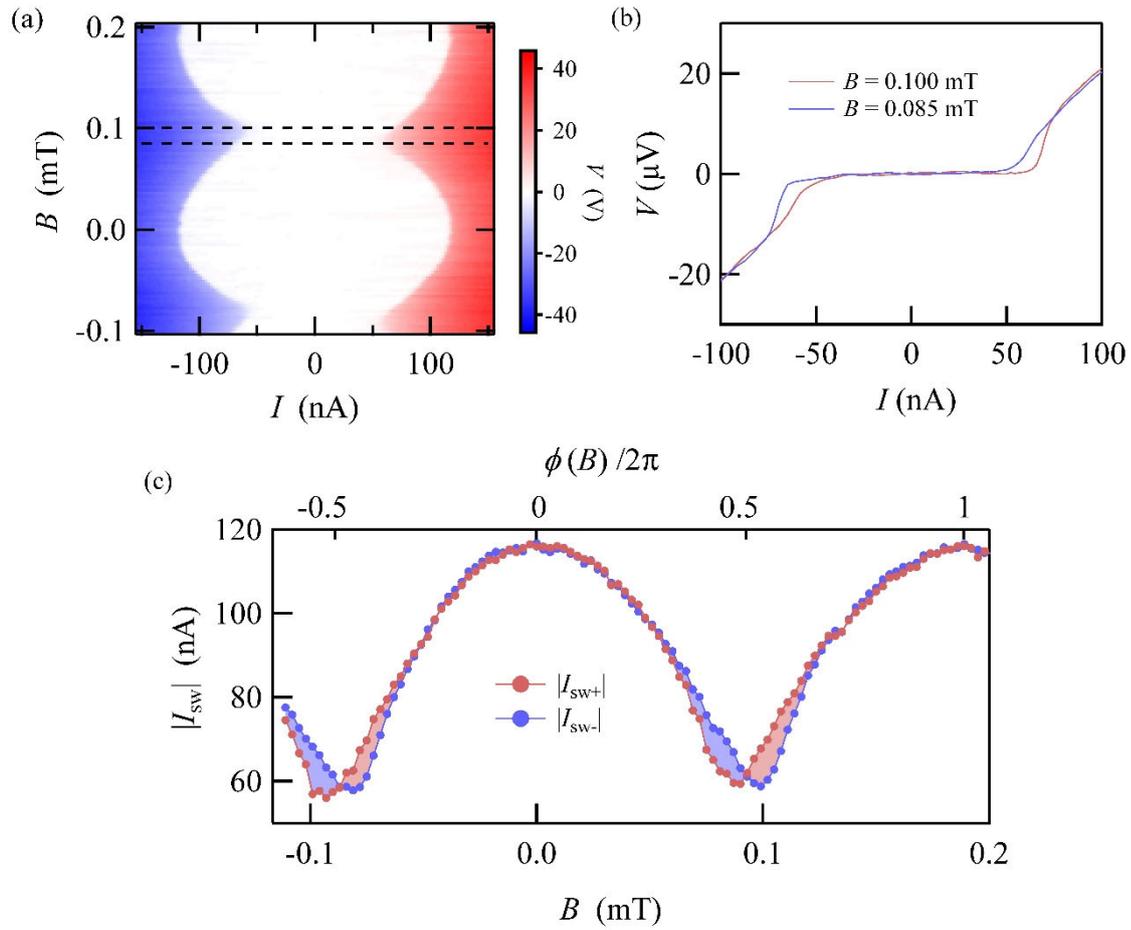

**Fig. S12**

(a) $V$ on JJU as a function of $I$ and $B$. The oscillation derived from the coherent coupling to JJL is clearly observed. (b) Line profiles of (a) for $B$ = 0.085 and 0.100 mT indicated by blue and red lines, respectively. The results indicate that the switching current in the positive $I$ sweep is different from that in the negative.

(c) $|I_{sw+}|$ and $|I_{sw-}|$ with respect to $B$ indicated by red and blue circles, respectively. It can be seen that a nonreciprocal supercurrent periodically emerges, as was observed in Fig. 2(c).

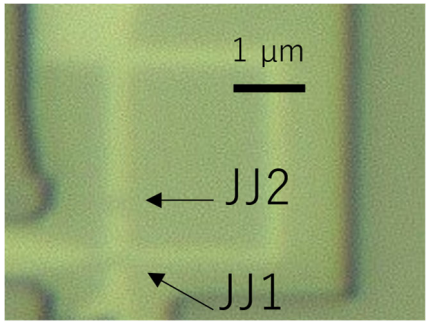 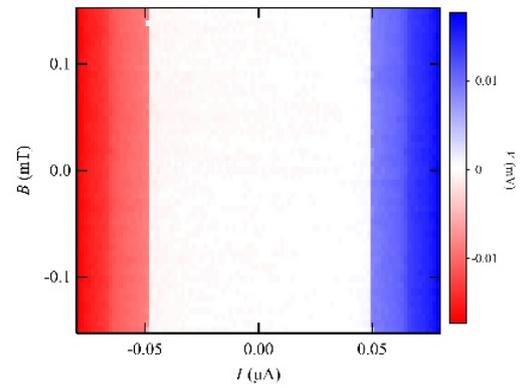

**Fig. S13**

(a) The optical image of the device with a wide separation of JJ1 and JJ2.

(b) The obtained voltage difference $V$ as a function of $B$ and $I$ which is the current flowing in JJ1. The switching current oscillation cannot be found.

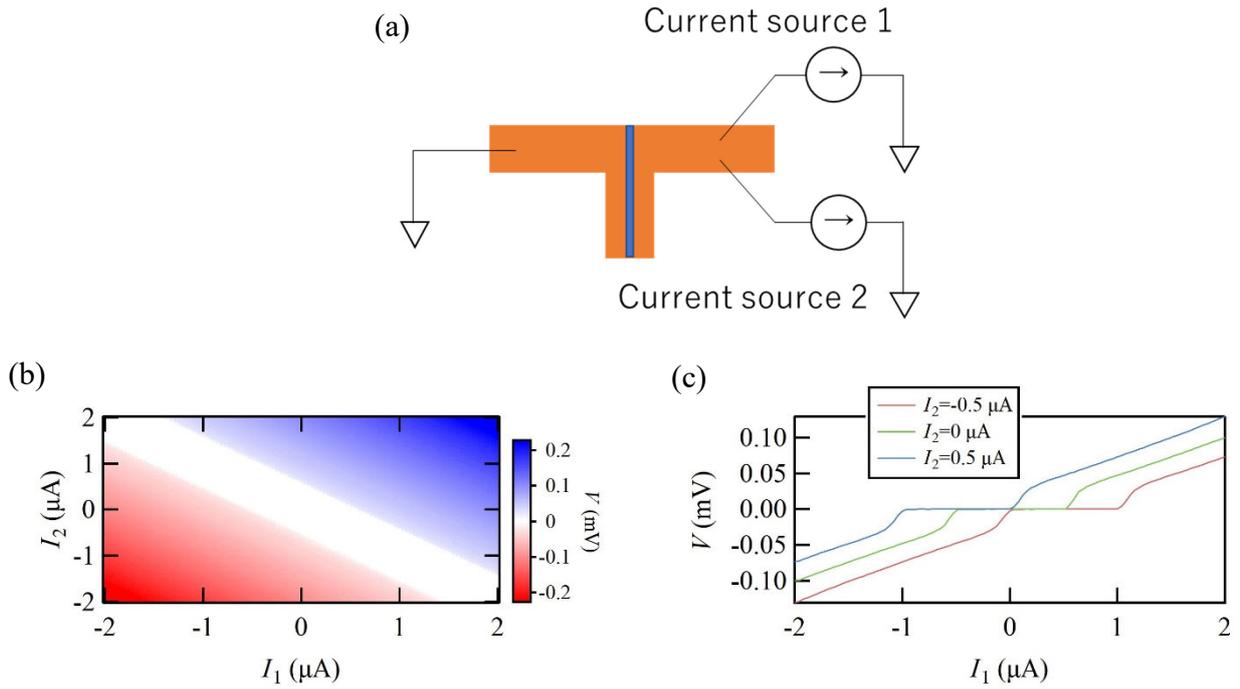

**Fig. S14**

(a) The schematic image of the measurement setup for the single JJ of Aluminium and the InAs quantum well with the two current sources. We define $I_1$ and $I_2$ as the bias currents from the current source 1 and 2, respectively.

(b) Voltage difference on the single JJ ($V$) as a function of $I_1$ and $I_2$.

(c) The line profiles of (b) at $I_2 = -0.5, 0, 0.5$ μA are shown as the red, green, and blue curves. With the finite $I_2$, the I-V curve can be shifted.